\documentclass[acmlarge]{acmart}

\usepackage{booktabs} % For formal tables
\usepackage{graphicx}
\usepackage{listings}
\usepackage{lipsum}
\usepackage{url}
\usepackage{multirow}
\usepackage{xcolor}
\usepackage[T1]{fontenc}
\usepackage[scaled]{beramono}
\newcommand\SSmall{\fontsize{9}{9.2}\selectfont}
\newcommand*\LSTfont{\SSmall\ttfamily\SetTracking{encoding=*}{-60}\lsstyle}

\newcommand{\CHANGED}[1]{#1}

\newenvironment{change}{\color{black}}{\color{black}}

\newenvironment{change2}{\color{black}}{\color{black}}

\newenvironment{change2b}{\color{black}}{\color{black}}

\usepackage[ruled]{algorithm2e} % For algorithms

\SetAlFnt{\small}
\SetAlCapFnt{\small}
\SetAlCapNameFnt{\small}
\SetAlCapHSkip{0pt}
\IncMargin{-\parindent}

\setcopyright{usgovmixed}
\acmDOI{0000001.0000001}

\received{January 2018}
\begin{document}
% Title portion
\title{Controlling Interactions with Libraries in Android Apps Through Runtime Enforcement}
%\titlenote{}

\author{Oliviero Riganelli}
\authornote{This is the corresponding author}
\affiliation{%
  \institution{University of Milano - Bicocca}
  \streetaddress{Viale Sarca 336}
  \city{Milan}
  \state{MI}
  \postcode{20126}
  \country{Italy}}
\email{oliviero.riganelli@unimib.it}
\author{Daniela Micucci}
\affiliation{%
  \institution{University of Milano - Bicocca}
  \streetaddress{Viale Sarca 336}
  \city{Milan}
  \state{MI}
  \postcode{20126}
  \country{Italy}}
\email{daniela.micucci@unimib.it}
\author{Leonardo Mariani}

\affiliation{%
  \institution{University of Milano - Bicocca}
  \streetaddress{Viale Sarca 336}
  \city{Milan}
  \state{MI}
  \postcode{20126}
  \country{Italy}}
\email{leonardo.mariani@unimib.it}

\begin{abstract}
Android applications are executed on smartphones equipped with a variety of resources that must be properly accessed and controlled, otherwise the correctness of the executions and the stability of the entire environment might be negatively affected. For example, apps must properly acquire, use, and release microphones, cameras, and other multimedia devices otherwise the behaviour of the apps that use the same resources might be compromised.  

Unfortunately, several apps do not use resources correctly, for instance due to faults and inaccurate design decisions. By interacting with these apps users may experience unexpected behaviours, which in turn may cause instability and sporadic failures, especially when resources are accessed. 

In this paper, we present an approach that lets users \emph{protect} their environment from the apps that use resources improperly by enforcing the correct usage protocol.  
This is achieved by using \emph{software enforcers} that can observe executions and change them when necessary. For instance, enforcers can detect that a resource has been acquired but not released, and automatically perform the release operation, thus giving the possibility to use that same resource to the other apps. 

The main idea is that software libraries, in particular the ones controlling access to resources, can be augmented with enforcers that can be activated and deactivated on demand by users to protect their environment from unwanted app behaviours. 
We call the software libraries augmented with one or more enforcers \emph{proactive libraries} because the activation of the enforcer decorates the library with \emph{proactive behaviours} that can guarantee the correctness of the execution despite the invocation of the operations implemented by the library. For example, enforcers can detect that a resource has not been released on time and proactively release it. 

Our experimental results with \begin{change}27 possible misuses of resources\end{change} in real Android apps reveal that proactive libraries are able to effectively correct \begin{change}library\end{change} misuses with negligible runtime overheads.

%
%Software libraries implement APIs that deliver reusable functionalities. To correctly use these functionalities, software applications must satisfy certain correctness policies, for instance policies about the order some API methods can be invoked and about the values that can be used for the parameters. If these policies are violated, applications may produce misbehaviors and failures at runtime. 
%
%Although this problem is general, applications that incorrectly use API methods are more frequent in certain contexts. For instance, Android provides a rich and rapidly evolving set of APIs that might be used incorrectly by app developers who often implement and publish faulty apps in the marketplaces. 
%
%To mitigate this problem, we introduce the novel notion of  \emph{proactive library}, which augments classic libraries with the capability of proactively detecting and healing misuses at runtime. Proactive libraries blend libraries with multiple proactive modules that collect data, check the correctness policies of the libraries, and heal executions as soon as the violation of a correctness policy is detected. The proactive modules can be activated or deactivated at runtime by the users and can be implemented without requiring any change to the original library and any knowledge about the applications that may use the library. 
%
%We evaluated proactive libraries in the context of the Android ecosystem. Results show that proactive libraries can automatically overcome several problems related to bad resource usage at the cost of a small overhead.
\end{abstract}



% DO NOT use this command unless you want to change
% the default behavior
% \authorsaddresses{Authors' addresses: G.~Zhou, Computer Science
%   Department, College of William and Mary, 104 Jameson Rd,
%   Williamsburg, PA 23185, US, \path{gzhou@wm.edu}; V.~B\'eranger,
%   Inria Paris-Rocquencourt, Rocquencourt, France; A.~Patel, Rajiv
%   Gandhi University, Rono-Hills, Doimukh, Arunachal Pradesh, India;
%   H.~Chan, Tsinghua University, 30 Shuangqing Rd, Haidian Qu, Beijing
%   Shi, China; T.~Yan, Eaton Innovation Center, Prague, Czech Republic;
%   T.~He, C.~Huang, J.~A.~Stankovic University of Virginia, School of
%   Engineering Charlottesville, VA 22903, USA; T. F. Abdelzaher,
%   (Current address) NASA Ames Research Center, Moffett Field,
%   California 94035.}

\maketitle

\renewcommand{\shortauthors}{O. Riganelli et al.}

\section{Introduction}\label{sec:introduction}

Users of mobile devices can easily and quickly access to a huge number of software applications through well-established digital marketplaces, such as the Android's Google Play\footnote{\url{https://play.google.com/store}}, the Apple's App Store\footnote{https://www.apple.com}, and the Microsoft's Windows Phone App Store\footnote{\url{https://www.microsoft.com/en-in/store/apps/}}. These marketplaces are freely populated with applications of various quality, ranging from professional apps to apps implemented by hobbyists. The openness of the marketplaces facilitates software distribution but also raises relevant reliability issues. In fact many of these apps are often not thoroughly validated and may cause troubles once installed and executed in a device.

%Software development for the mobile market has resulted in multiple ecosystems regularly accessed by a huge number of mobile device users. An ecosystem is typically composed of a marketplace (e.g., the Android's Google Play, the Apple's App Store, and the Microsoft's Windows Phone App Store), application providers (i.e., who implement and distribute apps through the marketplace), and customers (i.e., who download, install, and use apps available in the marketplace)~\cite{Hyrynsalmi:Ecosystems:IGIGlobal:2016}.

%The marketplace is open to everyone who likes to contribute by making apps available for download. However, the openness of such ecosystems raises reliability issues. In fact, the apps that are available on the marketplaces are not necessarily implemented by professional developers and might be unsafe and unstable. For instance, developers may %intentionally implement malicious apps that steal data from devices, or 
%accidentally implement and release faulty apps, which might cause issues in the devices where they are installed. 

To mitigate these problems, the marketplaces might be equipped with mechanisms to limit the proliferation of unreliable and insecure apps. For example, users can rate and review apps, so that unsafe and unreliable apps become less and less popular and fewer people try to install and use them~\cite{Kong:AUTOREB:CCS:2015}. Ecosystems may also implement static analysis routines to reject largely inadequate apps~\cite{Li:DroidRA:ISSTA:2016}. 

Although these mechanisms can be helpful to prevent the indiscriminate proliferation of undesired apps, customers regularly report issues with the apps downloaded from the marketplaces~\cite{Azim_Towards_2014,Banerjee:EnergyBugs:FSE:2014,Wei:AndroidFragmentation:ASE:2016,Shan:ResumerRestartErrors:OOPSLA:2016,Wu:Callback:TSE:2016}. A relevant portion of the problems experienced by the users is related to the way apps interact with the resources available in a device (e.g., the camera, the microphone, and the wifi antenna). For instance, many apps fail to properly acquire and release resources, causing efficiency and energy problems~\cite{Azim_Towards_2014,Banerjee:EnergyBugs:FSE:2014,Wu:Callback:TSE:2016}. In some cases, inaccurate interactions with the resources may even cause problems across apps. For instance, an app that does not release the camera every time its execution is suspended may prevent the other apps from acquiring the camera~\cite{Azim_Towards_2014,Wu:Callback:TSE:2016}.

Since many of these annoying problems can be addressed by forcing the faulty apps to satisfy the expected resource usage protocol, we studied how to design \emph{software enforcers} that can monitor executions and change the behaviour of the apps when necessary. To make the solution broadly applicable, software enforcers do not require access to the source code of the app, but work in a black-box fashion. Moreover, they could be speculatively activated or deactivated by users. For instance, enforcers might be activated after some applications presented some misbehaviours.  

%in this paper we present our work about giving to users the possibility to protect from apps that use resources improperly. In particular, we studied how users could enable and disable software enforcers that can monitor executions and change the behaviour of the apps when necessary. 

Interestingly, the incorrect interaction between an app and a resource can be often recognized by looking at both the usage of the library that controls the access to the resource and the lifecycle of \begin{change}the components of the app that interact with the library, for instance the lifecycle of an activity\end{change}\footnote{Android apps are composed of multiple components called activities: \url{https://developer.android.com/guide/components/activities}}. For \begin{change}example\end{change}, if the user moves to background an app that is using the microphone to record some audio, the app must immediately release the microphone, otherwise the other apps might be unable to interact with the microphone since it would be occupied by the inactive app, and the device may unnecessarily keep the microphone active consuming extra battery. At the level of the interaction between the app and the library this means that once an app has invoked the method \texttt{startRecording()} on the Android library class \texttt{AudioRecord}, it must release the microphone by invoking the method \texttt{release()} every time a call to the \texttt{onStop()} callback method is generated by the Android framework (\texttt{onStop()} is a callback method that apps implement to define the operations that must be performed before an activity becomes invisible to the user). Unfortunately, it is not always the case that apps are implemented releasing and acquiring resources coherently with the lifecycle of the activities~\cite{Liu:ResourceLeaks:ISSRE:2016,Azim_Towards_2014,Wu:Callback:TSE:2016}.

%Interestingly the incorrect interaction between an app and a resource \emph{can be often recognized by looking at the lifecycle of the activities\footnote{Android apps are composed of multiple components called activities.\newline\url{https://developer.android.com/guide/components/activities}} that compose the app and the calls to the API that controls the access to that resource}. For instance, if the user moves to background an app that is using the microphone to record some audio, the app must immediately release the microphone, otherwise \CHANGED{the other apps might be unable to interact with the microphone because it would be occupied by the inactive app, and the device may unnecessarily keep the microphone active consuming extra battery.} In the Android ecosystem, this means that once an app has invoked the method \texttt{startRecording()} on the \texttt{AudioRecord}, it must release the microphone by invoking the method \texttt{release()} every time a call to the \texttt{onStop()} callback method is generated by the Android framework (\texttt{onStop()} is a callback method that apps implement to define the operations that must be performed before an activity becomes invisible to the user). Unfortunately, it is not always the case that apps are disciplinely implemented releasing and acquiring resources coherently with the lifecycle of the activities~\cite{Liu:ResourceLeaks:ISSRE:2016,Azim_Towards_2014,Wu:Callback:TSE:2016}.

Enforcers can intercept interactions between apps and libraries to automatically fix these executions. Since enforcers are aware of the usage policies relevant to Android libraries, they are logically associated with libraries, that is, each library is decorated with a number of enforcers each one enforcing a different \emph{policy}. The presence of the enforcers turns regular libraries into \emph{proactive libraries}, which are libraries that can perform operations even when the methods of the libraries are not invoked (e.g., because the enforcer has to proactively execute operations to fix a detected problem). For example, when the \texttt{onStop()} callback is generated, the enforcer associated with the library for audio recording is triggered to check if the microphone has been released. If not, the enforcer may force the release of the microphone, and automatically reassign it to the activity once the activity becomes visible to the user again.

%To prevent users from experiencing annoying problems caused by apps that inaccurately interact with APIs, we introduce a novel concept of library that we called \emph{proactive library}. A proactive library is a library composed of two main parts: the reactive library, which is a regular implementation of an API, and a set of proactive modules, which decorate the library with the ability to enforce some \emph{correctness policies} at runtime, de facto augmenting the system with self-healing capabilities. 

%The proactive modules operate jointly with their library reacting to the invocation of certain API and callback methods, checking that the apps are using the API appropriately, and automatically fixing the execution and the status of the system, if necessary. 

Since a regular library does not need to be modified to be turned into a proactive library, enforcers can be easily added to existing libraries, without the need of designing libraries as proactive libraries from the very beginning. This eases the incremental adoption of the technology. Moreover, any party may contribute to the definition of the enforcers for a given library and not only the developers of the library. This aspect may facilitate the proliferation of the enforcers. \begin{change}App developers can keep using the usual testing and analysis methods to fix their apps. The design and implementation of enforcers is for library developers and in general for developers who want to contribute to the Android ecosystem, while the runtime activation of the enforcers is under the control of the final users.\end{change}

%LEO

%From an end-user perspective, the replacement of regular libraries with proactive libraries gives to users the opportunity to make their systems more stable, at the cost of a small overhead that might be introduced if many correctness policies are simultaneously violated. Moreover, the proactive modules can be turned on or off by users who can decide how to execute their applications case by case. This decision depends on the level of trusting on the installed apps and past experience. For instance, if users intend to use an app developed by an untrusted source or an app that caused problems in the past, they may activate the proactive modules before running the app, otherwise they may keep the proactive modules turned off and use the app normally.  

In addition to studying how to turn regular libraries into proactive libraries that can be exploited to make software systems more reliable, we defined a model-based approach for both the definition of the enforcers and the automatic generation of their implementation, that we call \emph{proactive module}, for the Android environment. Although the concepts of enforcers and proactive libraries are general, we studied them in the context of the Android ecosystem because many non-trivial and rapidly evolving APIs for resource management are available and problems with resources are frequent~\cite{Azim_Towards_2014,Banerjee:EnergyBugs:FSE:2014,Wu:Callback:TSE:2016}.

This work extends our previous work on proactive libraries~\cite{Riganelli:ProactiveLibraries:SEAMS:2017} by (i) introducing a process, a modelling environment, and a code generation strategy to automatically generate the proactive libraries from a specification of the enforcers, (ii) extending the modelling language used to specify the enforcers with special operations for restoring resources and with the possibility to distinguish if an action taken by an enforcer must be executed before or after an operation performed by the monitored software, (iii) increasing the sophistication of the proactive module to consider the status of \begin{change}Android components\end{change}, enforcement strategies, and resources simultaneously, (iv) providing a more detailed presentation of the approach and additional background material, (v) extending the empirical evaluation with \begin{change}15\end{change} additional real-world cases from popular Android markets and open-source Android community; \begin{change}and (vi) studying the scalability of the proactive modules and the usefulness of the model-based approach compared to the manual implementation of the enforcers.\end{change}

%\CHANGED{ experienced our approach on Android, }
%and focused on policies related to resource usage. 
%Although the concept of proactive library and correctness policy are general and not specific to the Android ecosystem, we think this solution is particularly relevant in the context of mobile systems where many non-trivial and rapidly evolving APIs for resource management are available. Our empirical evaluation based on policies for six different Android resources shows that proactive libraries can be a useful design approach to prevent problems at runtime.

%In a nutshell this paper provides the following contributions:
%\begin{itemize}
%\item It introduces the notion of \emph{proactive library}, which is a library decorated with proactive modules that can provide self-healing abilities for certain correctness policies,
%\item It presents a \emph{formalism} for the specification of the enforcement strategies,
%\item It describes an \emph{architecture} for the implementation of the proactive modules,
%\item It defines a process to \emph{automatically generate} proactive libraries for a target environment from their specification,
%\item It provides \emph{empirical evidence} about the effectiveness and efficiencies of proactive libraries.
%\end{itemize}

The paper is organized as follows. Section~\ref{sec:background} provides background information about edit automata, which is the formalism used in this paper to specify enforcers, model-driven software development, which is the methodology used to specify and generate the proactive libraries, \begin{change}and Xposed, which is the framework we used to implement the proactive modules for Android.\end{change} Section~\ref{sec:example} introduces a running example that is used throughout the paper to exemplify the approach. Section~\ref{sec:approach} presents the concept of proactive library and the process for implementing it. Section~\ref{sec:policy} describes how to formally define the correctness policies that can be enforced with proactive libraries. Section~\ref{sec:generation} presents how to automatically generate proactive libraries from their specifications. Section~\ref{sec:evaluation} presents our empirical evaluation. Section~\ref{sec:related} discusses related work. Section~\ref{sec:conclusion}  provides final remarks.

%Basically we foresee the possibility to use a library in two distinct ways: the classic reactive way and the proactive mode. In the classic reactive way the library responds to the requests produced by the application assuming that the application uses the library correctly, efficiently and effectively. In the proactive mode, the library is not only reactive, but also proactive, executing additional operations when the methods of the library are invoked and when callbacks are produced, checking the coherence between the status of the app and the status of the library, and fixing the situation if necessary. 

%The library is completely independent on the practice module, thus the mobile phone users will have the freedom to decide whether to execute apps, and thus libraries, in the normal model, or self-healed mode, which exploit the proactive module. In this way, users, depending on their level of trusting on the apps they execute and the problems experienced in the past, may decide whether to activate these additional routines.

\section{Background}\label{sec:background}
This section provides background information about edit automata, which is the formalism we used to model the behaviour of the enforcers, the model-driven software development paradigm, which is the development methodology we used to define the modelling environment and the generators of the enforcers, \begin{change}and the Xposed framework, which is the framework we used to implement the proactive modules\end{change}.

\subsection*{Edit Automata}
Enforcers are software components that monitor other programs to check their correctness and modify their behaviour when problems are detected. We model the behaviour of the enforcers using edit automata, which are abstract machines that can specify how a sequence of operations executed by a monitored program must be transformed into an alternative sequence of operations~\cite{Ligatti:Edit:JIS:2005}. In practice, we use edit automata to specify how to enforce behaviours that the monitored program may fail to satisfy.  

%Edit automata are mainly used to enforce behaviors that the monitored program may fail to satisfy. Here we use edit automata to enforce correctness policies.

More formally, an edit automaton is a finite state model with the capability to modify an input sequence of operations into an output sequence of operations by incrementally suppressing and inserting operations. \begin{change}Contrary\end{change} to a normal finite state machine, the transitions of edit automata are annotated with both an input operation and an output operation sequence. According to the semantics of the transitions, when the input operation is detected, the output sequence is emitted. The output sequence might be \emph{the same} \begin{change}as\end{change} the input operation, which corresponds to not altering the execution, may contain \emph{additional} operations, which corresponds to inserting additional operations in the execution, or may \emph{omit} the input operation, which corresponds to suppressing the input operation. %Finally, the output sequence may both suppress and insert operations. 

\begin{figure}[ht!]
\begin{center}
  \includegraphics[width=7cm]{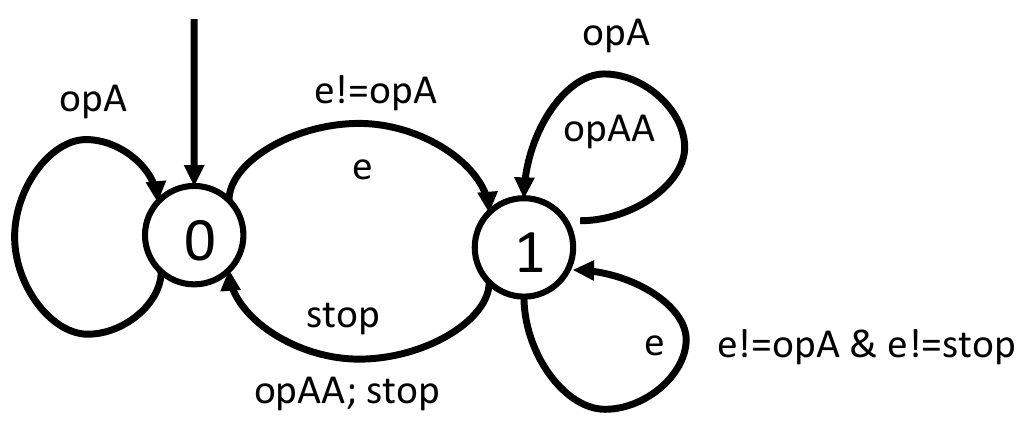}
\caption{A sample edit automaton.}
\label{fig:editAutomaton}
\end{center}
\end{figure}

Figure~\ref{fig:editAutomaton} shows a sample edit automaton. The symbol above each transition indicates the input operation recognized by the transition, while the sequence below a transition indicates the output sequence emitted by the automaton when the input sequence is recognized.
The input symbol can also be expressed with guard conditions, for instance \texttt{e!=opA} indicates any operation different from \texttt{opA}. \begin{change}In this case \texttt{e} is like a variable whose value depends on the concrete input operation that is recognized by the transition. If the event  \texttt{e} is also reported below the transition, the same event recognized in input is emitted in output\end{change}. The arrow pointing at state 0 indicates the initial state.

The sample edit automaton suppresses every execution of \texttt{opA} that is initially recognized (see the self-loop on state $0$). Once an event different than \texttt{opA} is detected, it moves to state $1$ where it leaves the execution unaltered unless \texttt{opA} or \texttt{stop} are detected (see the lower self-loop on state $1$). The occurrences of the operation \texttt{opA} are replaced with the occurrences of the operation \texttt{opAA} (see the upper self-loop on state $1$). While when a \texttt{stop} operation is recognized, the edit automaton introduces an extra \texttt{opAA} operation before the \texttt{stop} operation and goes back to state $0$.        

%In the sample edit automaton, the self-loop on state $0$ replaces every occurrence of the input event \texttt{doA} with the event \texttt{doX} (\texttt{doA} is suppressed and  \texttt{doX} is inserted) until an event different than \texttt{doA} is observed. When an event different than \texttt{doA} is observed at state $0$, the transition from state $0$ to state $1$ is taken and the model emits the input symbol as output (the input symbol is unmodified). At this point the self-loop on state $1$ replaces every occurrence of \texttt{doA} with the sequence of events \texttt{doA} followed by \texttt{doX} (\texttt{doX} is inserted). When an event different than \texttt{doA} is observed at state $1$, the transition from state $1$ to state $0$ is taken and the model emits the input symbol as output (the input symbol is unmodified).

For example, the edit automaton in Figure~\ref{fig:editAutomaton} transforms the input sequence

\smallskip
\noindent \texttt{opA}; \texttt{opA}; \texttt{opB}; \texttt{opA}; \texttt{opA}; \texttt{opC}; \texttt{stop} 

\smallskip
into the sequence 

\smallskip
\noindent  \texttt{opB}; \texttt{opAA}; \texttt{opAA}; \texttt{opC}; \texttt{opAA}; \texttt{stop}
\smallskip

When the edit automaton is used to specify how to enforce a policy for a given library, the input and output operations are events that correspond to method calls and to the termination of methods execution. These methods can belong to either the target library or \begin{change}to an object\end{change} lifecycle \begin{change}(e.g., an activity)\end{change}.

\subsection*{Model-Driven Software Development}
Model-Driven Software Development (MDSD) is \begin{change}an\end{change} approach for developing software by extensively exploiting models~\cite{Stahl:MDSD:2006}. MDSD \begin{change}extends the notion of\end{change} models as a form of documentation, \begin{change}to\end{change} consider them as artefacts equivalent to code, since they can be used to automatically generate executable code.

Code can be generated from models by applying a series of transformations, which can be organized in two logical sets: the implementation independent and the implementation dependent transformations. The implementation independent transformations are applied to an abstract model of the system to obtain a more concrete model that can be used as basis for code generation by applying the \begin{change}implementation\end{change} dependent transformations.

The abstract model expresses the semantics of the system regardless of the software platform used for its implementation (e.g., Android, iOS,  .NET, and so on). In general, the abstract model is at the same abstraction level of classic analysis models but represents the behavior of the system extensively. %in a nearly complete way. This property is important because the model must be extensive enough to be used for code generation.

To apply MDSD a number of tools are necessary, including modeling, model validation, and code generation tools. The modeling tools offer editors for the definition of the models using a domain specific modeling language based, for example, on UML profiles. Model validation tools verify the correctness of a given model with respect to a model domain-specific metamodel, which can be defined in a language such as OCL~\cite{OCL}. Code generation tools transform models into executable code from a given model. Code generation can only work if the model is correct with respect to the metamodel and therefore model validation is performed before code generation.

In our approach to the design of software enforcers, we implemented modeling tools to specify the enforcers, and defined both domain independent and domain dependent transformations to automatically generate the proactive modules that implement the behavior specified by the enforcers from the models.

\begin{change}
\subsection*{Xposed}
Xposed is a framework that can be installed in a rooted smartphone to allow users easily add modules that can customize the features present in the smartphone~\cite{Xposed_2016}. There exists a number of standard modules ready to be downloaded and installed. 

In addition to using existing modules, it is possible to implement new modules while exploiting the core capabilities of the Xposed framework, which include the ability to capture and modify the execution flow of Android apps. This ability is achieved by intercepting class loading events. Every time a class is loaded, its code can be instrumented on the fly before it is executed. The implementation of the module must specify the code locations that must be instrumented and the code that must be inserted at these places, similarly to what is done with aspect oriented programming~\cite{AOP97} frameworks. We exploit this framework to implement the software enforcers as Xposed modules.
\end{change}

\section{Running Example}\label{sec:example}
In this section, we present a motivating example based on an app for capturing pictures in Android, \textit{Plumeria}, that we downloaded from git (\url{https://github.com/DonLiangGit/Plumeria}). \begin{change}We selected this app because it is simple, but yet complex enough to illustrate the key concepts related to proactive libraries.\end{change}   

%\lstinputlisting[label=code, caption={A code-snippet of \textit{accentCheck}}]{MainActivity.java}

Listing~\ref{code} shows a code-snippet of \texttt{MyActivity}, which is the Android activity that uses the \texttt{Camera} in Plumeria. On startup, the callback method \texttt{onCreate()} is executed and a new \texttt{Camera} object is created by method \texttt{getCamera()}, whose implementation acquires the backward camera of the device invoking the method\linebreak \texttt{Camera.open()}. The method \texttt{Camera.open()} is an Android library method that can be used to request the access to the camera, while the corresponding library method for releasing the camera is \texttt{Camera.release()}.

\begin{change}The public list of issues for this app reports\end{change} an incorrect use of the camera\footnote{See issue \url{https://github.com/DonLiangGit/Plumeria/issues/1}.}. In fact, every time \texttt{MyActivity} becomes invisible, the camera becomes inaccessible to the other apps of the device. The camera is not even anymore accessible from Plumeria because when \texttt{MyActivity} is visible again, the \texttt{onCreate()} callback method is executed, and the call to \texttt{Camera.open()} produces an exception, since the app is attempting to acquire a camera that is already in use (by the app itself).

\begin{lstlisting}[label=code, caption={A code-snippet that shows the activity that handles the camera in Plumeria}]
...
public class MyActivity extends TabActivity {
    
    private Camera camInstance;
    ...
            
    protected void onCreate(Bundle savedInstanceState) {
        super.onCreate(savedInstanceState);
        setContentView(R.layout.activity_my);
        
        camInstance = getCamera();
        cameraPreview = new CameraPreview(this, camInstance);
        preview = (FrameLayout) findViewById(R.id.camera_preview);
        preview.addView(cameraPreview);
        ...
    }
    
    public static Camera getCamera() {
        Camera cam = null;
        cam = Camera.open();
        return cam;
    }
    
    ...
    @Override
    public void onPause() {
        super.onPause();
    }
    ...
}
\end{lstlisting}

This fault is caused by the app that does not handle the access to the camera properly. In particular, the \texttt{onPause()} callback method, which is invoked when the activity becomes invisible, does not invoke\linebreak \texttt{Camera.release()}, \begin{change}contrary\end{change} to what is recommended in the Android API documentation\footnote{See \url{https://developer.android.com/guide/topics/media/camera.html\#release-camera}.}.

%
%In fact, the \texttt{Camera.release()} should be  invoked every time the \texttt{MyActivity} is no longer visible to the user. 
%
%This requires invoking the  method \texttt{Camera.release()} inside the  the \texttt{onStop()} callback method. Since this app does not release the camera in the \texttt{onPause()}  callback method, every time the \texttt{MyActivity} becomes invisible, the camera would be unaccessible from all the other apps and the app itself. On the contrary, as recommended in the Android API documentation, the \texttt{Camera} should always be released when the \texttt{onPause()} method is called.

In the rest of the paper, we exploit this faulty app to present the concept of \begin{change}a\end{change} proactive library and to show how the proactive module associated with the \texttt{Camera} API may automatically detect and heal this problem.

\section{Proactive Libraries}\label{sec:approach}

A proactive library is a \emph{reactive library} augmented with one or more \emph{proactive module}s. The reactive library is a regular library that can be normally used as part of an application. In this paper we focus on Android libraries that control access to resources but the concept is generally valid for any library. The proactive modules are software components that enforce \emph{correctness policies} by altering executions, for instance by adding or suppressing method invocations. 

Although the modules are reactive per se, that is, they affect the execution when specific events are detected, their behaviour is proactive with respect to the associated library. In fact, the proactive modules can modify the executions regardless of the behavior of the library. In particular, proactive modules react to the behaviour of apps and libraries. The behaviour of the apps is observed in terms of the callback methods that are executed.  Invocations to \emph{callback methods} are calls produced by the framework (in this case the Android framework) to methods implemented in the app for correctly handling lifecycle state transitions. For example, in Android there are methods that are automatically invoked by the framework when an activity is suspended, destroyed, resumed, and so on~\cite{Android:Lifecycle:website}. 

The behavior of the libraries is observed by observing calls to \emph{library methods}, that is, calls to the methods implemented by the reactive library.

\begin{figure*}[ht!]
\begin{center}
  \includegraphics[width=13cm]{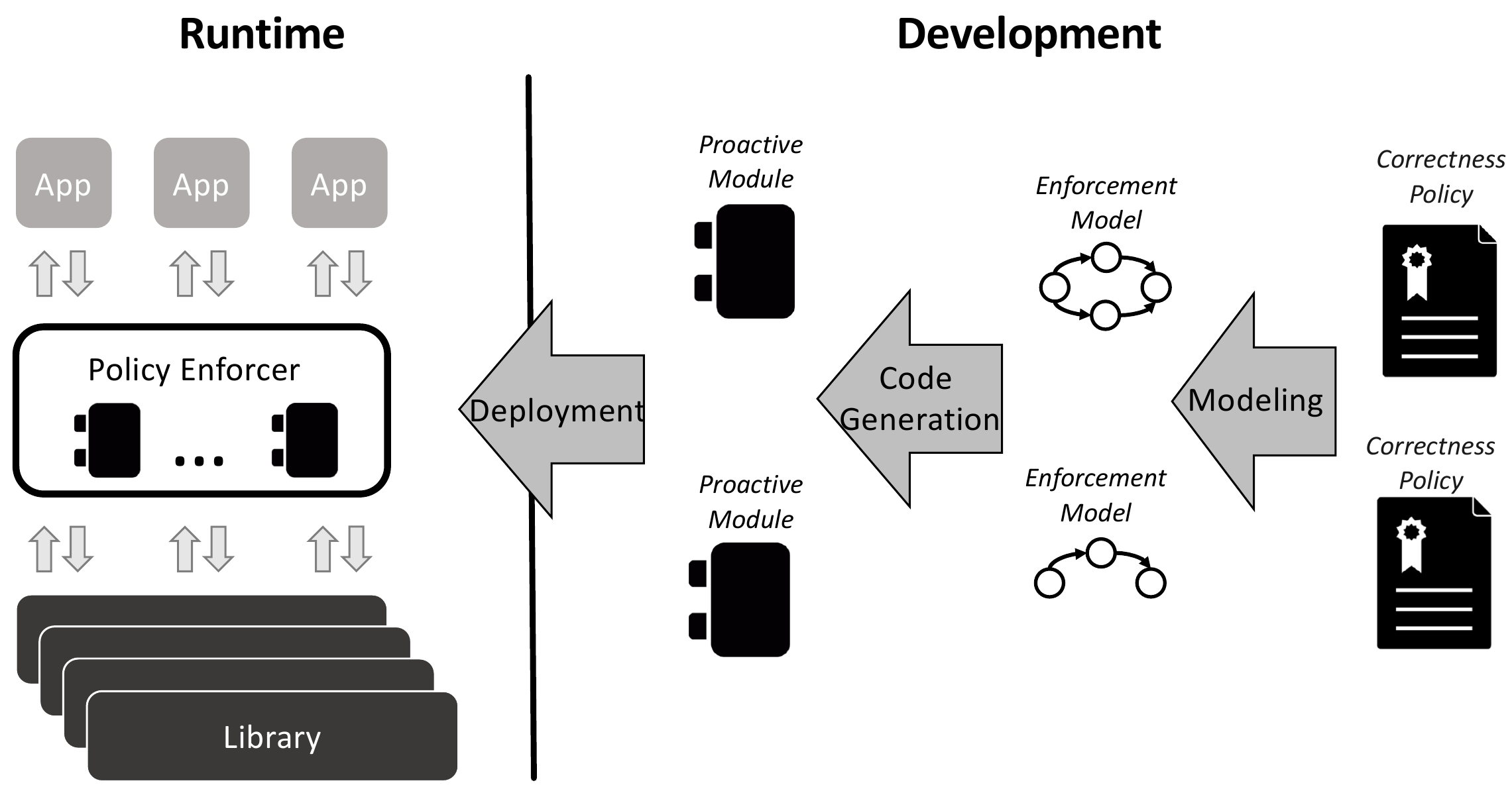}
\caption{Proactive Libraries.}
\label{fig:proactivelibrary}
\end{center}
\end{figure*}

Correctness policies specify how a library should be used by an app relating the status of the app, traced by observing calls to callback methods, to the status of the library, traced by observing calls to library methods. Correctness policies may \begin{change}potentially\end{change} cover any aspect related to the usage of a library, including functional correctness, privacy, and security aspects. In this paper we focus on a specific class of correctness policies that we call \emph{resource usage policies}. These are policies that state how the API interface of a library that controls the access to a resource should be used to prevent any misuse of the resource\CHANGED{, which may cause misbehaviors, failures and crashes at runtime}.

%Thus intuitively a proactive module can verify that a library is coherently by an app based on the used with respect to the current status of the app traced by intercepting the execution of the callback methods.

Note that in contrast to a regular reactive library, a proactive library may perform operations even when the library is not in use. For instance, the invocation of a callback method may trigger the execution of a proactive module, which may proactively perform some operations necessary to guarantee the correctness of the execution, regardless the direct involvement of the library in the computation. Considering the running example, a proactive module can fix the execution by forcing the release of the camera when \texttt{MyActivity} is no longer visible, that is, when the \texttt{onPause()} callback method is invoked, and re-assigning the camera to the app when the activity is resumed, that is, when the \texttt{onRestart()} callback method is invoked.

%guarantee the correctness of the execution even when the reactive library is not used by the app. In practice, the objective of the proactive modules is to relate the status of the apps, traced by intercepting the execution of callback methods~\cite{Android:Lifecycle:website}, to the status of the library, traced by intercepting the execution of the library API methods, and dynamically fix any detected issue.       

Figure~\ref{fig:proactivelibrary} visually illustrates how proactive libraries work, distinguishing the development and the runtime phases. At development time, developers of proactive modules first identify the correctness policies that they want to enforce. A \emph{correctness policy} is a natural language statement that constrains the usage of the API, considering the status of the app \begin{change2}as captured by its lifecycle\end{change2}, if necessary. Any app that violates a correctness policy is faulty. 

%\pagebreak
Examples of correctness policies, and more in particular resource usage policies since they refer to access to resources, for the \texttt{Camera} Android library are:

%\begin{enumerate}
%\item ``An app that becomes inactive while having the control of the microphone must first release the microphone.''
%\item ``An app cannot acquire the microphone twice.''
%\end{enumerate}
%Whenever an Activity of an app is stopped (its \texttt{onStop()} method is called), the \texttt{release()} method should be invoked to release the \texttt{AudioRecord} object.

\begin{quote}Resource Usage Policy 1: ``\emph{An activity that is paused while having the control of the camera must first release the camera}.''\end{quote}
\begin{quote}Resource Usage Policy 2: ``\emph{An app cannot acquire the camera twice}.''\end{quote} 

The first policy refers to both the status of the activity (an activity that is paused) and the usage of the API, while the second policy only refers to the usage of the API. Both policies are valid for every app that may use the camera. Note that the first policy is the one that should be exploited to heal the failure caused by the fault in the running example.

Correctness policies are then formalized and encoded as enforcement models. An \emph{enforcement model} precisely defines how to react to any violation of the correctness policies. We use edit automata~\cite{Ligatti:Edit:JIS:2005} to define the enforcement models because they represent a simple finite-state based formalization that naturally supports the specification of the behavior of the proactive module in terms of the events that must be intercepted and the events that must be inserted and suppressed as a reaction to each intercepted event. The objective of the enforcement model is to \emph{enforce} a correctness policy, when the running apps do not satisfy it. The enforcement model is defined uniquely using the knowledge of the library API and the Android lifecycle events (i.e., the callback methods)~\cite{Android:Lifecycle:website}, which are the same for any app. Thus its definition does not require any knowledge specific to the app that uses the library. See Section~\ref{sec:policy} for details about how to define enforcement models using edit automata. 

An enforcement model fully describes the behavior of the corresponding proactive module, which can be thus generated from the model. Proactive modules can be deployed in any environment where the corresponding library is used. Since proactive modules are activated by the invocation of certain methods, their execution in the user environment is controlled by a \emph{policy enforcer} that intercepts the events and dispatches them to the deployed proactive modules. The policy enforcer also controls the activation and deactivation of the proactive modules, which can be turned off and on by the user. See Section~\ref{sec:generation} for details about how to generate and execute a proactive module. 

\section{Enforcement Models}\label{sec:policy}

An enforcement model is a formal representation of the actions that must be undertaken to automatically enforce a correctness policy, that is, the operations that must be executed to turn an execution that violates a policy into an execution that satisfies it. In this paper, we focus on the Android ecosystem and on a specific class of correctness policies, the \emph{resource usage policies}, which state how an API that controls the access to a resource must be used by apps. An interesting aspect about these policies is that the usage of a resource is strongly coupled with the status of apps and their components, that is, depending on the state of an Android component \begin{change2}as captured by its lifecycle\end{change2} there are operations that must or must not be executed. 

In practice, an enforcement model traces the state of the \begin{change}app\end{change} by intercepting the system callbacks, \begin{change}for instance the system callbacks can be used to\end{change} uniquely identify the state of an activity according to the Android activity lifecycle~\cite{Android:Lifecycle:website,Riganelli:EnforcerReusability:ISOLA:2018}, traces the state of the library by intercepting the calls to the library API methods, and specifies the actions that must be proactively and automatically suppressed or inserted to enforce the satisfaction of a possibly violated policy. The proactive module obtained from the enforcement model guarantees the satisfaction of the policy at runtime, so that executions can be healed automatically.

In the following, we discuss how edit automata can be used to specify enforcement models.

\subsection*{Policy Specification}

An enforcement model formally defines the actions that must be automatically undertaken to enforce policies. In the case of resource usage policies for the Android ecosystem, the enforcement models can mainly suppress and insert actions of two kinds: callback methods and API methods. Callback methods are method calls produced by the Android framework when an app changes its status. For instance, the callback method \texttt{onStop()} is automatically invoked when the running activity of the app is no longer visible, while \texttt{onDestroy()} is invoked when the activity is destroyed. %Note that the set of callback methods is defined by the Android framework and does not depend on the specific app. 
The API methods are the methods implemented by the library associated with the proactive module. In the running example, the methods \texttt{open()} and \texttt{release()} are API methods implemented by the \texttt{Camera}. 

\begin{figure*}[ht!]
\begin{center}
  \includegraphics[width=10cm]{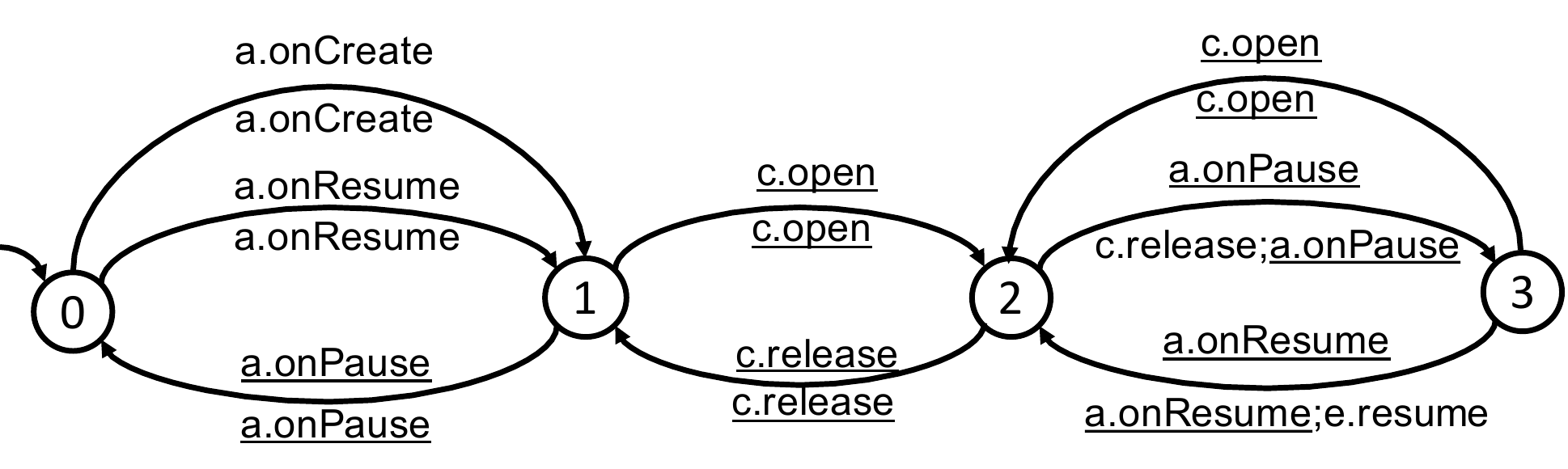}
\caption{Enforcement model for the Resource Usage Policy 1 of the \texttt{Camera}.}
\label{fig:simplifiedmodel}
\end{center}
\end{figure*}

Figure~\ref{fig:simplifiedmodel} shows the enforcement model for the Resource Policy 1 of the \texttt{Camera} API reported in Section~\ref{sec:approach}. Compared to general edit automata, our enforcement models extend the standard notation to support three aspects specific to our application context: prefixes, typed events, and special operations. 

We use \emph{prefixes} to disambiguate the element that performs the operation represented in the model. In this example, the prefix \texttt{a} represents \begin{change}the Activity class\end{change},  the prefix \texttt{e} \begin{change}represents the class that implements auxiliary methods that can be invoked by the enforcer\end{change}, and the prefix \texttt{c} \begin{change}represents the camera API. In the actual automata editor the full name of classes is used as prefix of methods\end{change}. 

The events in the model could be of two types: events that identify the beginning of an operation and events that identify the end of an operation. The type of an event can be distinguished based on its name: underlined event names correspond to the end of an operation, while regular names correspond to the beginning of an operation. For example, the outgoing transitions starting from state $0$ in Figure~\ref{fig:simplifiedmodel} represent the beginning of the execution of the callback methods \texttt{onCreate} and \texttt{onResume}, while the transition from state $1$ to state $2$ represents the completion of the execution of method \texttt{open} implemented by the camera.

This distinction is important to precisely represent the events that must activate the enforcers. For instance, the enforcer must capture the beginning of \texttt{onCreate} and not its end to \begin{change}check if\end{change} \texttt{open} is correctly invoked during the execution of this callback method, that is, after \texttt{onCreate} begins its execution but before it finishes it. Similarly, the enforcer must capture the conclusion of the callback method \texttt{onPause} from state $2$ to detect the case the resource has not been released, even while executing \texttt{onPause}.

Finally, the model may include events that are neither callback methods nor API methods but whose semantics is known by the code generator. This capability is exploited to address cases that can be hardly managed within the enforcer but that can be easily addressed with appropriate tool support. These events are represented with the  prefix \texttt{e}, to indicate enforcer methods, such as the event \texttt{e.resume} in Figure~\ref{fig:simplifiedmodel}. 

In this case, the event is used to activate a resume policy specific to the target resource. The policy could be in principle explicitly encoded in the model, but this would unnecessarily complicate the specification of the enforcers. To address these cases, we rather extended the code generator with special operations of known implementation that can be embedded in the generated code, as discussed in Section~\ref{sec:generation}. In the context considered in this paper, we used this capability only to support resume operations, since they may require the generation of multiple code fragments that trace the values of the variables used in the implementation of the resume procedure. 

\smallskip

The example model in Figure~\ref{fig:simplifiedmodel} enforces the release of the \texttt{Camera} when the activity goes to background while it still has the camera. In the initial state (state $0$), the model waits for the creation of a new activity. Once the activity has been created (state $1$), the enforcer checks whether a camera is opened.  If a camera is opened, the enforcer checks that it is released before the activity is paused (transitions from state $2$ to state $1$ and from state $1$ to state $0$). In all these cases the execution is unaltered because the app would be using the camera consistently with the policy that requires the camera to be released every time the activity is suspended.

However, the app might be paused without releasing the camera (transition from state $2$ to state $3$). In this case the enforcer changes the execution forcing the release of the camera. When the execution of the activity is recovered, the camera is reassigned back to the activity \begin{change}if the activity does not already take the camera back autonomously\end{change} (transitions from state $3$ to state $2$), so that the behavior of the enforcer is fully transparent to the app.

Using edit automata to specify the behavior of a proactive module compared to directly coding the module has several advantages: (1) it provides a compact representation of the enforcement strategy that can be more easily manipulated and modified than working at the code level, (2) it can be used to automatically generate an implementation of the enforcement strategy, reducing the risk of introducing faults and speeding up the implementation process, and (3) it can be composed with other finite state specifications (e.g., the specification of other policies, the app lifecycle, and the protocol of the API) to formally check its correctness~\cite{RV17}. 

\section{Generation of Proactive Modules}\label{sec:generation}

A proactive module is a software component that verifies the correctness of the execution according to the strategy defined in an enforcement model. Since implementing proactive modules manually is an error prone and time consuming process, we defined a Model-Driven Software Development (MDSD) process and the corresponding tool chain to obtain proactive modules automatically from the enforcement models.

%The development of a software component that verifies the correct execution of a program with respect to a set of constraints and that forces the execution in case of violation of constraints, passes from a sequence of activities that goes from the modeling of the edit automaton to the deployment of the corresponding proactive module.
%These activities can be performed manually by a developer, but are time-consuming, require experience from the developer, and are error-prone.
%
%Therefore, to help developers in building proactive modules, we have developed an environment based on Model-Driven Software Development that facilitates the modelling and validation of edit automata and the generation and deployment of the corresponding proactive modules.

%we build a comprehensive tool that allows developers to apply the MDSD approach for the development of proactive modules to be deployed on the target platforms. The aim of the tool is to ease the modelling and validation activities of edit automata and the generation and deployment of the corresponding proactive modules in the target platforms. %and to make the generation of code automatic with respect to the target platform. 

%Following the MDSD approach, the environment supporting the developer in generating proactive modules should include support for the following activities. 
Our MDSD tool chain implements the core functionalities of a MDSD environment:
\begin{itemize}
\item \emph{Modelling}, which provides the capability to model enforcement models that encode resource usage policies;
\item \emph{Validation}, which provides the capability to validate the syntactic correctness of an enforcement model;
\item \emph{Code generation}, which provides the capability to automatically generate the source code that implements the enforcement logic described by the enforcement model; 
\item \emph{Compilation} and \emph{deployment}, which provide the capability to compile and deploy the generated proactive module on a target platform.
\end{itemize}

We assigned these responsibilities to two main components: the \emph{EMEditor}, which implements \emph{modelling} and \emph{validation} capabilities, and the \emph{PMGenerator}, which implements \emph{code generation}, \emph{compilation}, and \emph{deployment} capabilities.

\begin{figure*}[ht!]
\begin{center}
  \includegraphics[width=15cm]{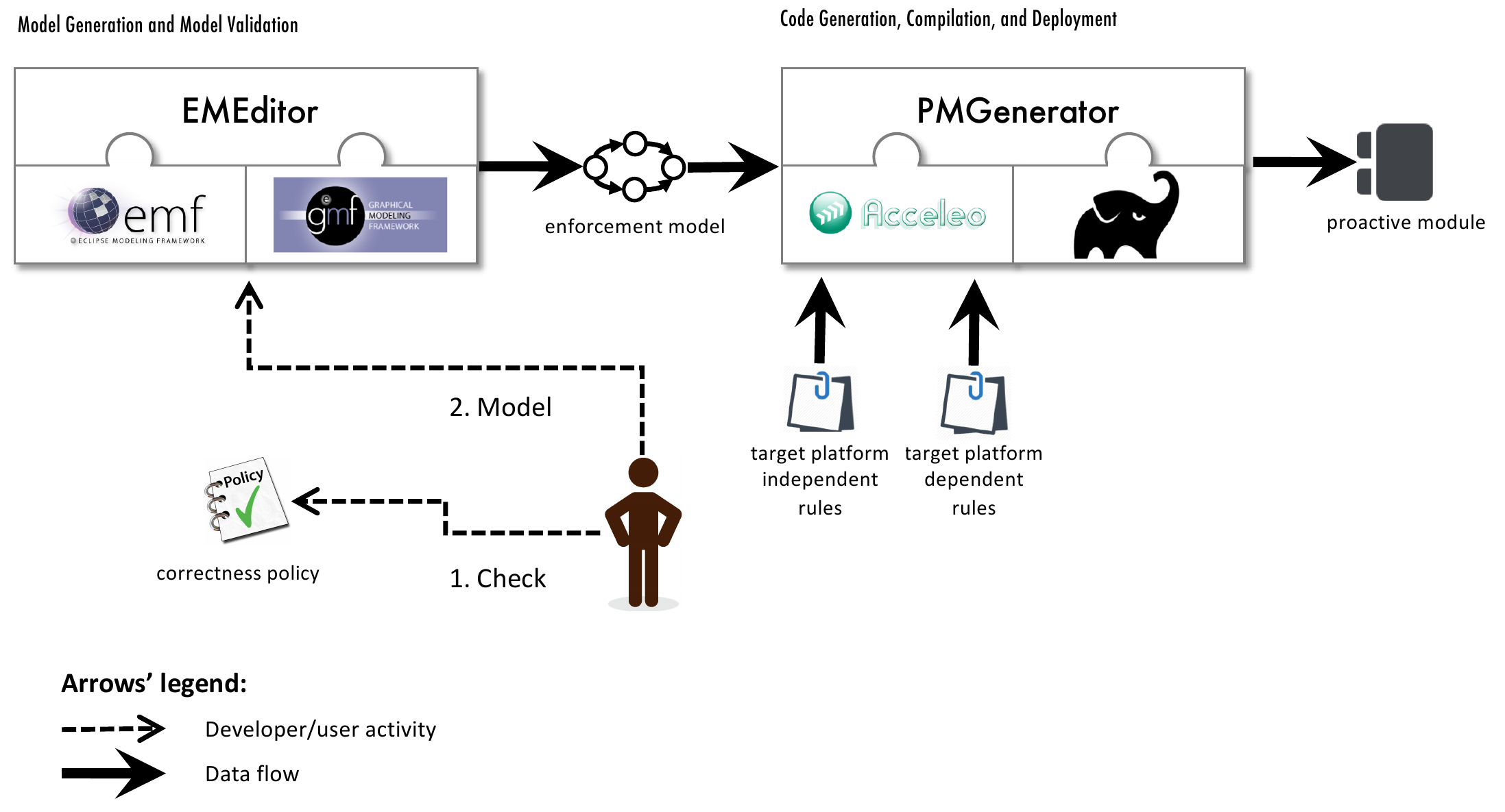}
\caption{The end-to-end tool chain of our MDSD-based environment}
\label{fig:tool}
\end{center}
\end{figure*}

Figure~\ref{fig:tool} shows the activities that a developer who uses our MDSD-based environment follows to define enforcement models and generate the proactive modules. The developer uses the \emph{EMEditor} to visually design a model that enforces the considered policy. The output of the EMEditor is a validated enforcement model, which is the starting point for the generation of the proactive module. The developer then uses the \emph{PMGenerator} to generate and deploy the proactive module corresponding to the input enforcement model. Code generation is based on two sets of rules that distinguish the platform independent and the platform dependent part of the process.
%takes as input the edit automaton generated by the EMEditor and set of rules that drives the code generation, compilation, and deployment activities. The output is a proactive module to be deployed on the target platform.
In the following, we describe the EMEditor and the PMGenerator. 

\subsection{EMEditor}
\begin{figure*}%[ht!]
\begin{center}
  \includegraphics[width=10cm]{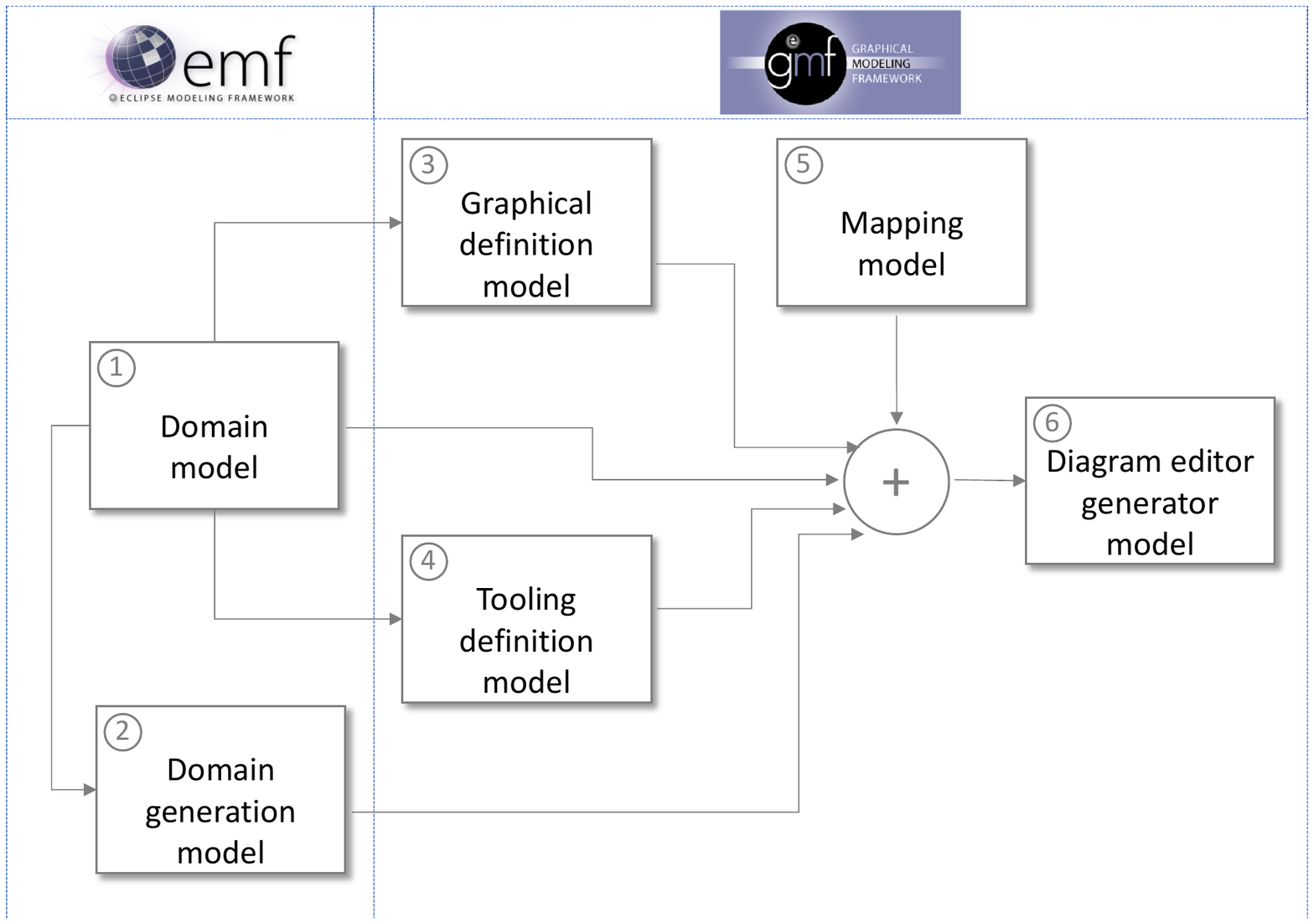}
\caption{The models and their composition that define the EMEditor}
\label{fig:EMEditor-models}
\end{center}
\end{figure*}

We implemented the EMEditor in Eclipse~\cite{Eclipse} using the Eclipse Modeling Framework (EMF)~\cite{EMF} and the Graphical Modeling Framework (GMF)~\cite{GMF} plugins. EMF can be used to create modeling and code generation tools starting from a data model. GMF can be used to implement graphical editors in Eclipse. 
We thus created the EMEditor by defining models in EMF and obtaining the corresponding graphical editors with GMF. Figure~\ref{fig:EMEditor-models} shows the set of models that we defined and combined. 

\begin{figure*}%[ht!]
\begin{center}
  \includegraphics[width=11cm]{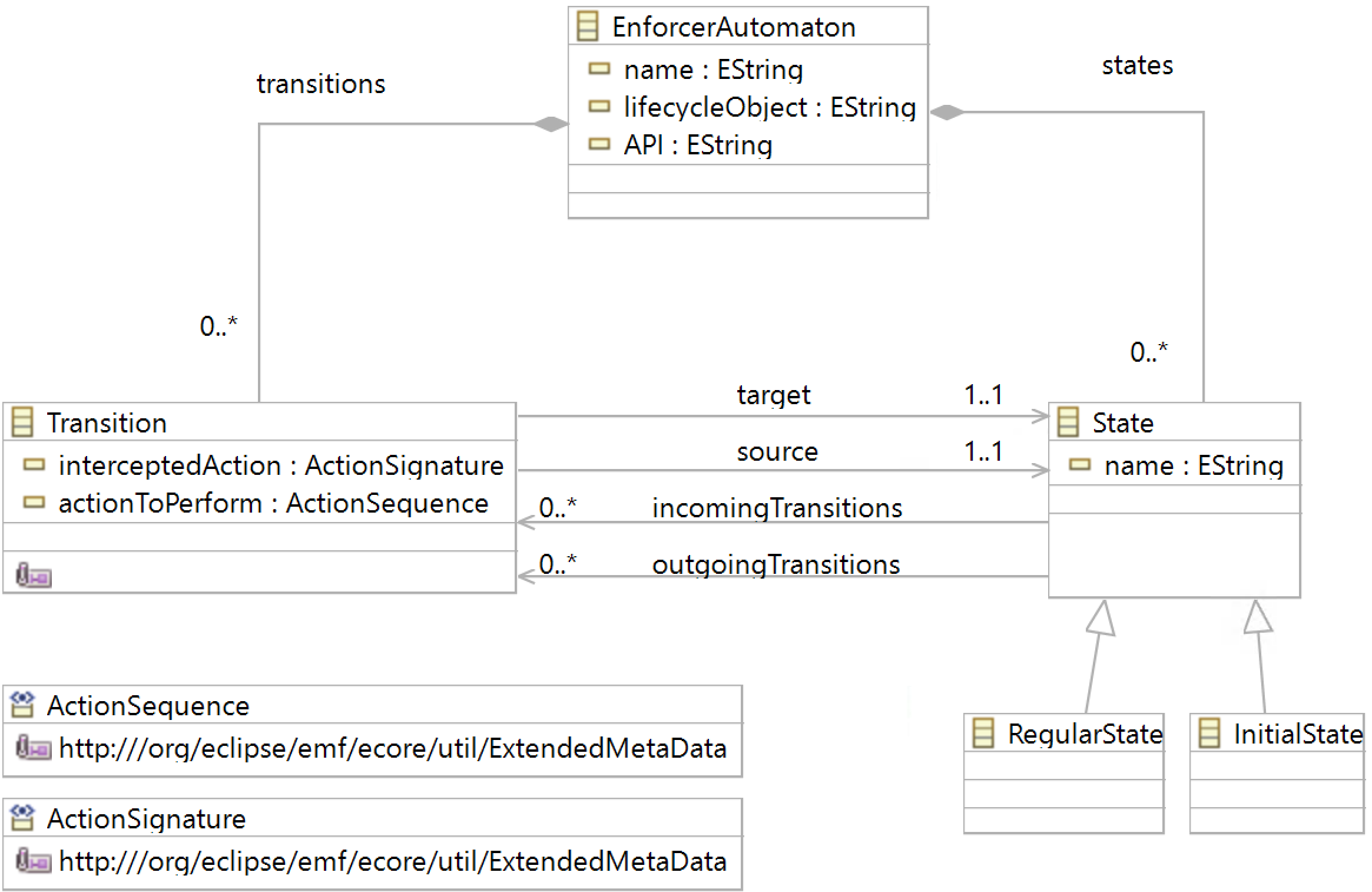}
\caption{The domain model of an enforcement model defined with the EMF framework}
\label{fig:EAModel}
\end{center}
\end{figure*}

The \emph{domain model} is the meta-model that defines what enforcement models are. When the developers use the EMEditor to define enforcement models they create instances of the classes defined in the domain model. Our domain model is shown in Figure~\ref{fig:EAModel}.

%We first defined the \emph{domain model}, that is the meta-model that describes an enforcement model, in the EMF framework.The domain model specifies the classes from which the EMEditor will allow to create instances. It is shown in Figure~\ref{fig:EAModel}. 

\begin{change}The \texttt{EnforcerAutomaton}  is characterized by a \texttt{name}, a \texttt{lifecycleObject} that uses the API, and the \texttt{API} that we want to be used correctly.  The automaton\end{change}
%that is characterized by a name (\texttt{name}); the name of the technology to be used to generate the corresponding proactive module (\texttt{technologyToUse}); the full name of the application to be enforced  (\texttt{targetApplicationFullQualifiedName}); the name of the package in which the source code of the proactive module will be inserted within the Android project (\texttt{sourcePackageName}); the list of full names of all the classes whose instantiations must be intercepted by the proactive module t(\texttt{classesToHookAllConstructors}). This allows the proactive module to keep track of all the resources whose usage has to be monitored and possibly enforced. 
is represented as a set of \texttt{States} connected by \texttt{Transitions}. A \texttt{State} could be an \texttt{InitialState} or a \texttt{RegularState}. \texttt{Transitions} are annotated with both the name of the operation that is intercepted, attribute \texttt{interceptedAction}, and the possibly empty sequence of actions that are generated by the enforcer, attribute \texttt{actionToPerform}. Those two attributes are associated with two type definitions that constrain the values of these strings according to a syntax that is checked by the editor. The regular expressions associated with type definitions are not shown in the model, however an \texttt{ActionSignature} is constrained to be a method name with a prefix that indicates whether the enforcer should intercept the execution of the method when it is invoked (prefix \texttt{before\#}) or when it returns (prefix \texttt{after\#}). The \texttt{ActionSequence} is constrained to be a comma separated sequence of action signatures.
 
We then derived the \emph{domain generation model} from the domain model through a standard EMF model-to-model transformation. This model enriches the domain model with additional information useful to the editor, such as, the set of properties associated with each graphical element. 

%In principle, EMF is enough to automatically obtain an editor for modeling the instances of the defined domain model. However, the resulting editor is \textbf{TBD}

%Using EMF it is possible to obtain, through the model-driven generation, an editor useful for modeling instances of the defined domain model.  However, this editor is tree-based, allowing to graphically represent a model instance only through a hierarchical representation (where each model entity is represented by a predefined icon accompanied by a label that displays the name). 

%This results in offering to developers a poor and cumbersome interaction. 
To obtain the visual part of the EMEditor we exploited GMF, which can be used to generate an editor  where each entity of the model has a user-defined visualization.
% making the definition and the management of the model instances simpler, faster, and more intuitive. 
%
To define the editor with GMF we defined four models, as shown in Figure~\ref{fig:EMEditor-models}.

The \emph{graphical definition model} associates a graphical representation with each element of the domain model that must be displayed within the drawing sheet of the editor. In our case, we defined shapes for all kinds of states and transitions. %Figure~\ref{fig:EMEditor-graphicalDefinitionModel} shows an example related to the states of the enforcement model. Each geometric shape is described by means of an element called \emph{figure descriptor}, which specifies the dimension that the shape must have, the list of geometric forms of which it is composed, and the layout with which to organize them.

%\begin{figure*}[ht!]
%\begin{center}
%  \includegraphics[width=15cm]{EMEditor-graphicalDefinitionModel}
%\caption{The geometric shapes associated to the initial, regular, and final sates, and, for each of them, its description in the \emph{graphical definition model}}
%\label{fig:EMEditor-graphicalDefinitionModel}
%\end{center}
%\end{figure*}

The \emph{tooling definition model} defines the palette and the other accessory tools, such as menus and toolbars. In our case, we defined a palette with four tools: three tools to create states of two types (initial and regular) and one tool to create transitions. We also included features provided by GMF, such as the ability to zoom in and out and the possibility to add notes. %Figure~\ref{fig:EMEditor-palette} shows the palette for the creation of the states and the transitions. As shown in Figure~\ref{fig:EMEditor-palette}, the palette is defined by an element of \emph{palette type} in which it is possible to define the tool group element that contains the list of all the tools included in the palette.

%\begin{figure*}[ht!]
%\begin{center}
%  \includegraphics[width=8cm]{EMEditor-palette}
%\caption{The palette of the editor containing the tools to add a new state (initial, regular, or final) and a new transition}
%\label{fig:EMEditor-palette}
%\end{center}
%\end{figure*}

The \emph{mapping model} defines how to obtain the components of the editor, that is, the \emph{diagram editor generator model}, by combining the domain model, the graphical definition model, the tooling definition model, and the domain generation model. In practice the mapping model mainly associates the elements in the domain model (i.e., states and transitions) with both the respective geometric shapes defined in the graphical definition model and the palette tools defined in the tooling definition model. %In addition, the attributes of the \texttt{State} and \texttt{Transition} elements have been associated with the relative labels declared in the geometric shapes contained in the graphical definition model, to establish the value that must be displayed by each label. 
The constraints defined in the domain model (e.g., the constraints on the annotations associated with transitions) are also propagated to the diagram editor generator model. The mapping model includes additional constraints %written with OCL 
that verify the integrity of the model drawn in the editor (e.g., the presence of an initial state and the existence of the Java methods specified in the model).

 %\texttt{self.sates -> forAll(s | s.incomingTransitions -> size() > 0 or outgoingTransitions -> size() > 0}). 

%Figure~\ref{fig:EMEditor-validation} shows all the defined constraints  to be checked when validating an edit automaton. The set of constraints is specified by declaring an \emph{audit container} type element, containing an \emph{audit rule} element for each constraint to be defined. A constraint is defined through the declaration of a \emph{constraint} element and of a \emph{domain target} element that defines on which class of the domain model the satisfaction of the constraint must be verified.
%
%\begin{figure*}[ht!]
%\begin{center}
%  \includegraphics[width=11cm]{EMEditor-validation}
%\caption{Definition of the constraints for the validation of an edit automaton}
%\label{fig:EMEditor-validation}
%\end{center}
%\end{figure*}

%inally, the \emph{diagram editor generator model} contains implementation details for the generation of the editor.

%Thanks to the presence of the mapping model, it is possible to reuse the models defined for the generation of the modelling tool. The mapping model, in fact, makes the domain model, the graphical definition model, and the tooling definition model interchangeable.

\begin{figure*}[ht!]
\begin{center}
  \includegraphics[width=\textwidth]{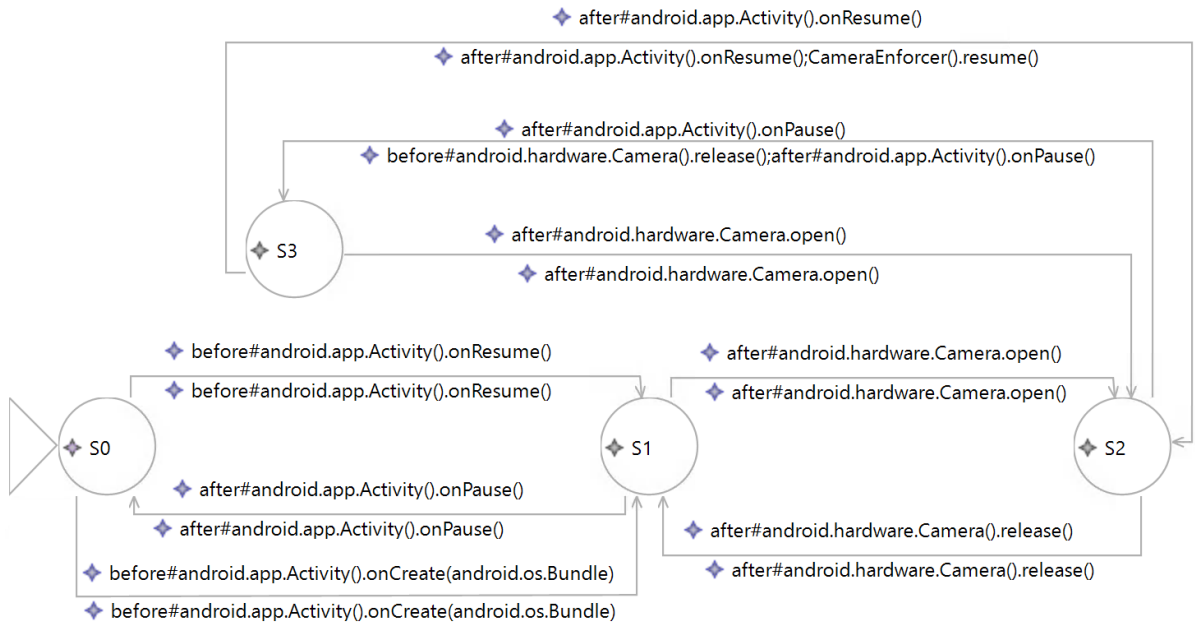}
\caption{The enforcer corresponding to the model in Figure~\ref{fig:simplifiedmodel} modeled with the EMEditor.}
\label{fig:runningExample-editor}
\end{center}
\end{figure*}

\begin{change}Finally\end{change}, the EMEditor is automatically obtained from the diagram editor generator model using GMF, \begin{change}as shown in Figure~\ref{fig:EMEditor-models}\end{change}.
%The resulting EMEditor can be used to specify enforcement models following an MDSD approach. 
Figure~\ref{fig:runningExample-editor} shows the enforcement model corresponding to the one defined in Figure~\ref{fig:simplifiedmodel} drawn in the EMEditor. The models can be saved as XML 
%files (XML Metadata Interchange)~\cite{XMI} 
which are then processed by the PMGenerator. % standard defined by the OMG (Object Management Group)~\cite{OMG}. 

%that we have specified using the EMEditor 
%
%has been generated using the EMEditor and corresponding to the one defined for the running example in Figure~\ref{fig:simplifiedmodel}. The events to be intercepted and to be emitted correspond to the beginning or the end of a method execution. Thus, to specify them, we used the fully qualified signature of the method including the formal parameters when present. We inserted in the fully qualified name of the method a pair of round parentheses between the class that defines the method and the method itself to specify that we are referring to an instance method. Moreover, to distinguish between the beginning or the end of a method execution, we used the prefixes ``before\#'' and ``after\#'' respectively. 
%For example, the label  \texttt{before\#android.app.Activity().onResume()} on transition from state \texttt{s0} to state \texttt{s1} specifies that the event to be intercepted is the beginning of the instance method \texttt{onResume} defined in the class \texttt{Activity}.

%\subsection{Architecture of a Proactive Module}

\subsection{PMGenerator}

The enforcement model defined in the EMEditor can be used to generate the actual proactive module that is deployed in the target environment to actuate the enforcement strategy. In this section, we first present the architecture of the proactive module, and then we describe how the components of a proactive module can be generated automatically. 

\medskip

We designed the architecture of a proactive module to be generic, relying only on the possibility to implement two features with the selected technologies in the target environment: (1) the ability to intercept method calls, and (2) the ability to encapsulate the instance of the library controlled by the proactive module. These two features can be obtained in many different ways with different technologies and in different environments. The information about the specific technologies to use is encoded in the sets of rules used by the PMGenerator to emit the proactive module.

\begin{figure*}[ht!]
\begin{center}
  \includegraphics[width=10cm]{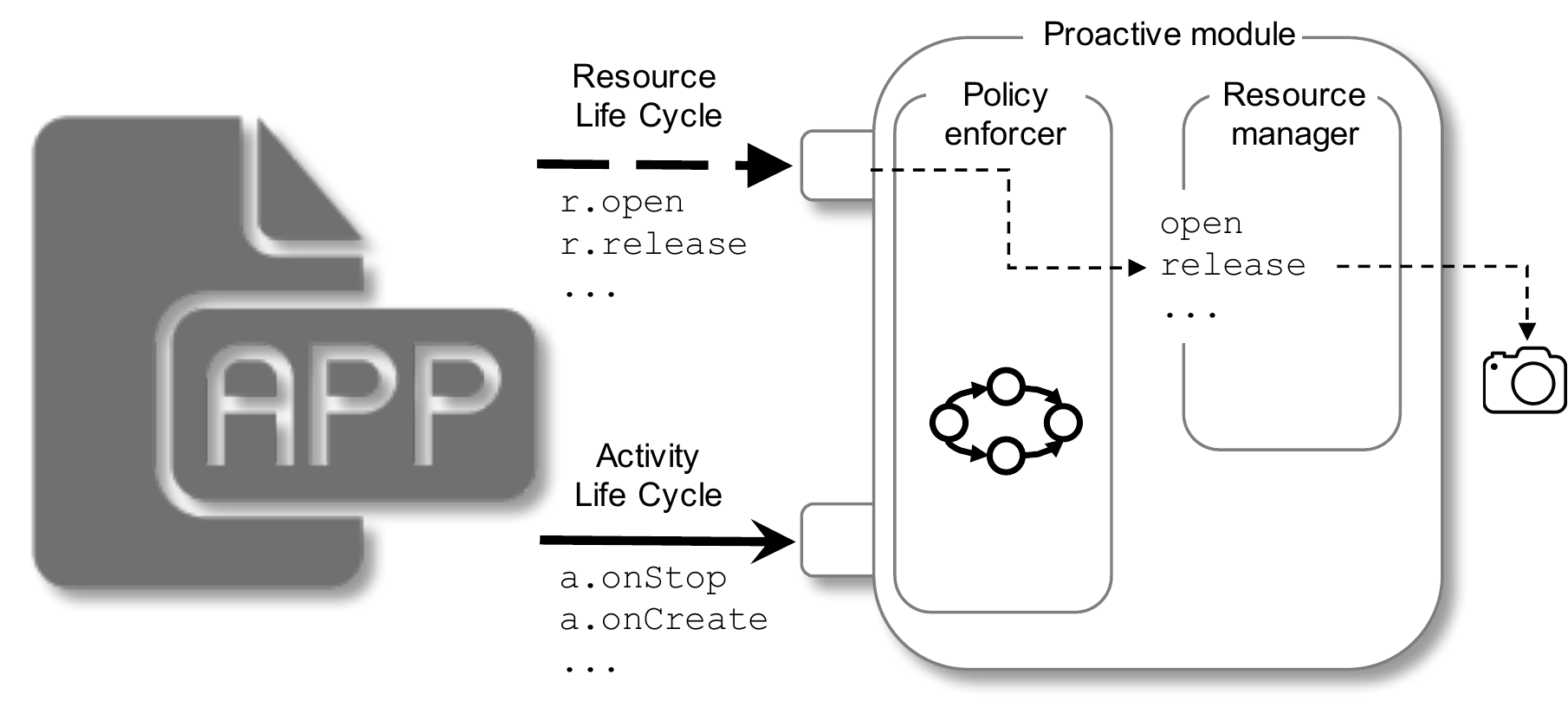}
\caption{The architecture of the proactive module}
\label{fig:enforcer}
\end{center}
\end{figure*}

To assess our approach, we implemented the PMGenerator for Android using Xposed~\cite{Xposed_2016}, which allows to cost-efficiently intercept method invocations and change the behavior of an Android app using run-time hooking and code injection mechanisms. Figure~\ref{fig:enforcer} graphically illustrates the architecture of a proactive module instantiated on Android. 

Xposed transparently intercepts the method calls to the encapsulated resource and to the object life cycles and propagates them to the policy enforcer, which reacts based on the behavior defined in the enforcement model. For instance, it might introduce additional method calls or suppress calls.

Every interaction with the target library, including interactions originated from the app and from the policy enforcer, is mediated by a resource manager. In this way, the reference to the target library is stored inside the resource manager and it can be destroyed and recreated transparently to the app. This capability is sometime relevant to the policy enforcer, for instance when there is an instance that must be destroyed and recreated to release and acquire a resource. \begin{change}Similarly, the resource manager detects the identity of the Android component that interacts with a given library instance, preventing the application of the enforcement strategies to cases not supported.\end{change} The resource manager is also injected in the application using Xposed to redirect method calls.

\medskip

%The goal of our tool-chain was to obtain a proactive module  with a structure like the one shown in Figure~\ref{fig:enforcer}. The ``policy enforcer'' module is in charge of intercepting the callback method related to the Android activity lifecycle and the method calls on the resource. 

%The intercepted callback methods of the Android activity are possibly allowed to be executed according to the enforcement rules. The intercepted methods on the resource are delegated to the ``resource manager'' module whose interface is exactly the one of the resource. The ``resource manager'' executes the intercepted methods on a private reference to the resource. 
%%This architecture makes sure that the callback methods of the Android activity are intercepted and let run according to the enforcement rules, while the methods on the resource are also intercepted, but executed on a private instance of the resource. 
%This allows the proactive module to make method calls to the resource even if the resource is private to the application to enforce.

%Thus, we select the Android Studio~\cite{AndroidStudio} framework for the compilation and deployment activities. 
%A proactive module responds to certain events occurred in the monitored app (e.g., callbacks and calls to library methods). Thus, its implementation requires to rely on a framework for intercepting and reacting to events. Since our MDSD-based environment has been implemented for Android apps, we selected the Xposed framework~\cite{Xposed_2016} that allows to cost-efficiently intercept method invocations and change the behavior of an Android app using run-time hooking and code injection mechanisms. To avoid rooting, alternative methods, such as bytecode instrumentation~\cite{Redexer_2017}, could be also exploited to implement the policy enforcers.

%The PMGenerator provides support for code generation compilation, and deployment of proactive modules. We used the Eclipse~\cite{Eclipse} framework and the Acceleo~\cite{Acceleo} plugin for the code generation. Acceleo is an open-source code generator that relies on the model-driven approach. Whereas, we used the Gradle~\cite{Gradle} build tool for compilation and deployment. Gradle is a tool that  allows developers to easily build and deliver software solutions.

To automatically generate the proactive module from the model defined with the EMEditor tool, we used  Eclipse~\cite{Eclipse} augmented with the Acceleo~\cite{Acceleo} plugin, which is an open-source code generator for model-driven development. 

As shown in Figure~\ref{fig:tool}, the code generation task uses two sets of rules that define how the resulting code should be structured, distinguishing between the rules that depend on the target platform (e.g., Android and Xposed) and the ones that are independent (e.g., rules that depend on the choice of the target language online).

\begin{figure*}[ht!]
\begin{center}
  \includegraphics[width=9cm]{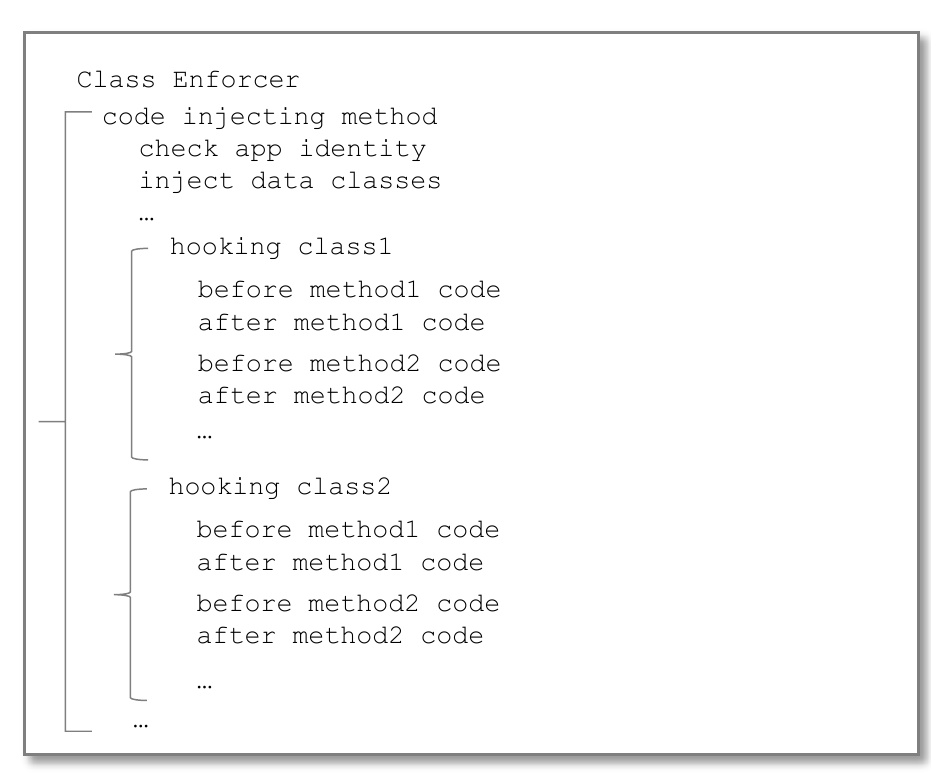}
\caption{Logical structure of the implementation of an enforcer for Android}
\label{fig:PMGenerator-rules}
\end{center}
\end{figure*}

The platform-independent rules map the \begin{change}elements\end{change} in the enforcement model to an implementation in a given target language. In our case, we use the switch-case design pattern~\cite{Kelly:CodeGeneration:2008} to properly organize the generated code in a Java class. The platform-dependent rules generate the additional code necessary to intercept method calls in the target environment (e.g., Android) with a given technology (e.g., Xposed). %and propagate these calls to the methods produced by the platform independent rules.

%are target platform independent and specifies how to codify states and transitions in terms of programming constructs. Among the state and the switch-case patterns, we selected the switch-case because it makes the source code easier to implement. In addition, the complexity of a enforcer model is not such that it justifies the use of a state pattern.

\begin{lstlisting}[float=!ht, label=listing1, caption={A code-snippet of the proactive module corresponding to the enforcement model in Figure~\ref{fig:simplifiedmodel}}]
public class CameraReleaseEnforcer extends IXposedHookLoadPackage {
   ...
   public void handleLoadPackage(XC_LoadPackage.LoadPackageParam lpparam) throws Throwable {
      ...
      Class<?> apiClass = findClass("android.hardware.Camera", lpparam.classLoader);
      Class<?> lifeCycleObjectClass = findClass("android.app.Activity", lpparam.classLoader);
      ArrayList<Classes<?>> appClasses = getAppClassesFromDex(lpparam);
      ...
      for (Class<?> appClass : appClasses) {
         ...
         if(apiClass.isAssignableFrom(appClass)){
            findAndHookMethod(appClass, "open", new XC_MethodHook() {
               ...
               protected void afterHookedMethod(MethodHookParam param) throws Throwable {
                  ...
	          <@\textcolor{blue}{\underline{//transition from state 1 to state 2 (camera open)}}@>
                  <@\textcolor{blue}{\underline{if (S1.equals(currentStates.get(lifeCycleObject)))} \{ }@>
                      <@\textcolor{blue}{\underline{currentStates.put(lifeCycleObject, S2);}}@>
                  <@\textcolor{blue}{\underline{\}}}@>
                  ...
               }
            });
            ...
         }
         ...
         if(lifeCycleObjectClass.isAssignableFrom(appClass)){
            findAndHookMethod(lifeCycleObjectClass, "onPause", new XC_MethodHook() {
               ...
               protected void afterHookedMethod(MethodHookParam param) throws Throwable {
                  ...
                  <@\textcolor{blue}{\underline{//transition from state 2 to 3 (activity on pause)} }@>
                  <@\textcolor{blue}{\underline{if (S2.equals(currentStates.get(lifeCycleObject))) }\{ }@>
	           <@\textcolor{blue}{\underline{doNotAlterateExecution = true;}}@>
                     Camera resource = lifeCycleObject2resource.get(lifeCycleObject);
                     <@\textcolor{blue}{\underline{saveCameraState(resource, lifeCycleObject);}	}@>
                     <@\textcolor{blue}{\underline{resource.release();}}@>
                     <@\textcolor{blue}{\underline{doNotAlterateExecution = false;}}@>
                     <@\textcolor{blue}{\underline{currentStates.put(lifeCycleObject, S3); } }@>
                  <@\textcolor{blue}{\underline{\}} }@>
                  ...
               }
            });
            ...
         }
      }
   }
}
\end{lstlisting}

Figure~\ref{fig:PMGenerator-rules} shows the high-level structure of the file generated automatically for Android using the Xposed framework. Each enforcer is a class with a single method (the \emph{code injecting method}) that implements the code injection logic. The method is invoked every time the device loads new packages. The initial part of the method (\emph{check app identity}) checks if the loaded package belongs to an app that must be affected by the enforcer, otherwise it returns immediately. The rest of the code (\emph{inject data classes}) injects auxiliary classes that might be used from the code injected afterward, for instance the implementation of the resume operation in the running example.

The code injecting method then includes a series of blocks that inject code in the individual classes that must be instrumented (\emph{hooking class} methods). Each block looks for the class to instrument first, and then injects specific behaviours that are executed before (\emph{before method code}) and after (\emph{after method code}) the target method is executed. The injected code fragments are themselves defined as methods that are executed before or after the target method.

Listing~\ref{listing1} shows an excerpt of the code corresponding to the model in Figure~\ref{fig:simplifiedmodel} generated by the PMGenerator. %with Xposed and Android as target platforms.
%From a technical perspective, Xposed can be used to specify the methods that must be executed as reaction to a specific event. 
The blue \begin{change}underlined\end{change} code is the code obtained with platform independent rules.

%The \texttt{handleLoadPackage()} method is the method responsible of instrumenting the loaded packaged. 
\begin{change}The \texttt{handleLoadPackage()} method is the method responsible for instrumenting the loaded packages. The invocations to \texttt{findClass} search for the classes that are specified in the model and that must be instrumented.
The \texttt{getAppClassesFromDex} extracts the classes in the application package. In the
cycle that iterates over these classes, the code checks if the class in the app package is the same or a subclass of the classes that have to be instrumented. If the class has to be instrumented, the \texttt{findAndHookMethod} invocation identifies and instruments the target methods. For instance, the sample code instruments the \texttt{open} method of the \texttt{android.hardware.Camera} class and its subclasses, and the \texttt{onPause} method of the \texttt{android.add.Activity} class and its subclasses.\end{change}
%The sequence of invocations \texttt{findClass} and \texttt{findAndHookMethod} looks for the class to be instrumented and instruments the target method, respectively. For instance, the sample code instruments the \texttt{open} method of the class \texttt{android.hardware.Camera} and the \texttt{onPause} method of the \texttt{android.add.Activity} class.

When a method is instrumented, the body of the method \texttt{findAndHookMethod} may contain the \linebreak\texttt{beforeHookedMethod} and the \texttt{afterHookedMethod} inner methods, which define the operations that must be executed before and after the target method, respectively. 

The structure of the enforcer discussed so far is the result of rules that depend on the target platform. The body of the \texttt{beforeHookedMethod} and \texttt{afterHookedMethod} however may derive  from rules of both kinds.

The code that depends on the platform is the one that retrieves and stores variables from data structures to make them accessible by the code generated from the enforcement model, such as the code that retrieves the current state of the enforcer (\texttt{currentStates.get(lifeCycleObject)}) and the code that retrieves the resource associated with the current activity (\texttt{resource = lifeCycleObject2resource.get(lifeCycleObject)}). Note that the implementation stores and retrieves this information while taking into account the mapping between the activity, the resource manipulated by the activity, and the enforcer that controls that interaction (a different enforcer instance is created for each pair activity-resource).

The rest of the code is generated from the enforcement model only. The code that derives directly from the model could be as simple as changing the current state only, such as for the \texttt{afterHookMethod} of the \texttt{open} method, or it might be more complex. For instance, the \texttt{afterHookMethod} associated with the \texttt{onPause} method generates an additional method call (\texttt{resource.release}) to force the release of the camera. The code also includes statements to avoid that the enforcer intercepts the calls produced by the enforcer itself (see usage of variable \texttt{doNotAlterExecution}) and code %, generated due to the presence of operations that require (\texttt{e.resume}), 
to store variable values to later enable the possibility to resume the resource.

Once the file has been generated, it can be compiled and deployed on the target platform. Our MDSD- environment exploits Gradle to import the generated source file in an Xposed project defined in Android Studio IDE~\cite{AndroidStudio}. \begin{change}Our prototype implementation of the tool can be downloaded from \url{https://gitlab.com/learnERC/proactivelibrary}\end{change}.

\begin{change}Note that the entire process is independent on the apps that will be affected by the enforcers. \begin{change2}In fact, our MDSD-environment does not require to know the identity of the apps that may violate a policy but only the library API involved in the policy that must be enforced.\end{change2} The target apps are identified by the user when enabling the enforcement mechanism for one or more target apps. The enforcement mechanism checks the identity of the app in the \emph{check app identity} step before the rest of the enforcement logic is applied. If the app is not among the ones to be enforced, its behavior is not modified.\end{change}

%Usando gradle, il file generato viene importato all'interno di un progetto XPosed utilizzando Android Studio (usando gradle?). Viene quindi compilato
%+ compilazione
%+ lancio

%At the aim we created once for all an empty structure with the Android Studio IDE that includes an XPosed module.
%creo una struttura 'vuota' con Android studio che includa il progetto di un modulo XPosed (una tantum). 

%con gradle, copio, compilo e lancio (build e deployment)

\section{Evaluation}\label{sec:evaluation}
To evaluate proactive libraries, we considered \begin{change}three\end{change} research questions, one about the effectiveness, one about efficiency\begin{change}, and one about the usefulness of the MDSD environment\end{change}.

\smallskip
\noindent \emph{RQ1 (effectiveness): Can proactive libraries be used to detect and heal actual library misuses in Android applications?}

With this research question, we evaluate the effectiveness of proactive libraries against actual problems reported in Android apps. As discussed in the paper, we focus on the misuses of the APIs that control the access to local resources because these misuses are quite frequent~\cite{Liu:ResourceLeaks:ISSRE:2016,Azim_Towards_2014,Wu:Callback:TSE:2016}, and may have a significant impact on the system, for instance affecting the other non-faulty apps that run in the same device and use the same resources.
\smallskip

\noindent \emph{RQ2 (efficiency): What is the overhead introduced by proactive libraries compared to traditional reactive libraries?}

With this research question, we measure the runtime overhead that can be experienced when the proactive modules are active compared to the standard configuration that executes the reactive libraries only. 
\smallskip

\begin{change}
\noindent \emph{RQ3 (usefulness): Can developers conveniently obtain proactive modules using the MDSD environment?}

With this research question, we compare the effort needed to obtain proactive modules using the MDSD environment that we defined to the effort needed to directly implement them in Java. 
\end{change} 
\smallskip

\begin{table}[]
\begin{change}
\caption{Apps}
\label{tab:apps}
\resizebox{16cm}{!}{%
  \centering
\begin{tabular}{lccccc}
\toprule
\textbf{App Name} & \begin{change2}\textbf{Version} \end{change2}& \textbf{Type} & \textbf{Downloads} & \textbf{Last Update} & \begin{change2}\textbf{Links}\end{change2}                                                                                                                                            \\ \midrule
AndroidHacks      &  \begin{change2}N/A \end{change2} & open source   & -                  & 2018                 & \begin{change2}https://github.com/YingVickyCao/AndroidHacks/        \end{change2}                                                                                                      \\ \midrule
BlueChat          &  \begin{change2}1.1 \end{change2}& open source   & -                  & 2016                 & \begin{change2}https://f-droid.org/wiki/page/com.alexkang.bluechat  \end{change2}                                                                                                      \\ \midrule
CheDengWo         &  \begin{change2}2.4.4 \end{change2}& closed source & \begin{change2}2.3 millions\end{change2}       & 2018                 & \begin{change2}https://www.qimai.cn/andapp/downTotal/appid/4676           \end{change2}                                                                                                \\ \midrule
ErWeiMaL          &  \begin{change2}1.0 \end{change2}& closed source & \begin{change2}4.2 millions\end{change2}       & 2015                 & \begin{change2}http://apps.wandoujia.com/apps/com.weilaifu.qrcode  \end{change2}                                                                                                       \\ \midrule
FontMaster        &  \begin{change2}4.3.9 \end{change2}& closed source & \begin{change2}894 millions \end{change2}      & 2018                 & \begin{tabular}[c]{@{}c@{}}\begin{change2}https://www.wandoujia.com/apps/com.zhiqupk.ziti\end{change2}\\ \begin{change2}https://www.qimai.cn/andapp/downTotal/appid/607\end{change2}\end{tabular}                  \\ \midrule
Foocam            &  \begin{change2}1.2 \end{change2}& open source   & \begin{change2}100+  \end{change2}             & 2017                 & \begin{change2}https://play.google.com/store/apps/details?id=net.phunehehe.foocam2           \end{change2}                                                                             \\ \midrule
FromCat           &  \begin{change2}580 \end{change2}& closed source & \begin{change2}20,000\end{change2}             & 2016                 & \begin{change2}https://www.qimai.cn/andapp/downTotal/appid/143359     \end{change2}                                                                                                    \\ \midrule
GetBackGPS       &  \begin{change2}0.4.1 \end{change2} & open source   & -                  & 2019                 & \begin{change2}https://ruleant.github.io/getback\_gps/   \end{change2}                                                                                                                 \\ \midrule
IPST2          &  \begin{change2}N/A \end{change2}   & open source   & -                  & 2017                 & \begin{change2}https://github.com/fRueD4096/IPST2         \end{change2}                                                                                                                \\ \midrule
MaMa              &  \begin{change2}1.0 \end{change2}& closed source & \begin{change2}10,000+ \end{change2}          & 2015                 & \begin{change2}https://www.qimai.cn/andapp/downTotal/appid/881807       \end{change2}                                                                                                  \\ \midrule
QiCaiScan         &  \begin{change2}1.0 \end{change2}& closed source & \begin{change2}19,300+ \end{change2}           & 2018                 & \begin{change2}https://www.qimai.cn/andapp/downTotal/appid/151774       \end{change2}                                                                                                  \\ \midrule
SuperTorch        &  \begin{change2}3.2 \end{change2}& closed source & \begin{change2}17.9 millions \end{change2}     & 2015                 & \begin{tabular}[c]{@{}c@{}}\begin{change2}https://www.wandoujia.com/apps/com.idiantech.torch\end{change2}\\ \begin{change2}https://www.qimai.cn/andapp/downTotal/appid/25771\end{change2}\end{tabular}             \\ \midrule
WebPCsuite        &  \begin{change2}3.1.6 \end{change2}& closed source & \begin{change2}1.5 millions  \end{change2}     & 2016                 & \begin{tabular}[c]{@{}c@{}}\begin{change2}https://www.appsapk.com/web-pc-suite/\end{change2}\\ \begin{change2}https://apk-dl.com/web-pc-suite-file-transfer/com.geeksoft.wps\end{change2}\end{tabular}             \\ \midrule
WiFiSaver         &  \begin{change2}N/A \end{change2}& open source   & -                  & 2018                 & \begin{change2}https://github.com/FilipBanak/WifiSaver        \end{change2}                                                                                                            \\ \midrule
XiaoMiWifi        &  \begin{change2}1.1.755 \end{change2}& closed source & \begin{change2}32.9 millions\end{change2}      & 2018                 & \begin{tabular}[c]{@{}c@{}}\begin{change2}https://play.google.com/store/apps/details?id=com.xiaomi.router\end{change2}\\ \begin{change2}https://www.qimai.cn/andapp/downTotal/appid/1463\end{change2}\end{tabular} \\ \bottomrule
\end{tabular}
}
\end{change}
\end{table}

To investigate these \begin{change}three\end{change} research questions, we selected a set of Android apps that have been reported to be affected by API misuses. In particular, we selected all the apps from the benchmark that Liu et al.~\cite{Liu:ResourceLeaks:ISSRE:2016} used to evaluate their static analysis technique for the detection of resource leaks. This consists of 9 closed-source apps and 3 open-source apps. \begin{change}We extended this set with 3 open source apps from GitHub\end{change} for a total of 15  real-world Android apps. \begin{change}Table~\ref{tab:apps} lists all the apps, reporting their name (column \emph{App Name}), \begin{change2}version name when available (column \emph{Version})\end{change2}, the type of app (column \emph{Type}), the number of downloads of the app for the ones available on a store (column \emph{Downloads}), the year last time the app has been updated either on the open source repository or on its online store (column \emph{Last Update}), \begin{change2}and the link to download the app from either the open source repository or the app store (column \emph{Links})\end{change2}.\end{change} The apps are from a variety of domains, including communication and productivity, \begin{change}are quite popular, and still maintained\end{change}. All these apps use resources and have been reported to be potentially affected by a total of \begin{change}27\end{change} resource misuses spanning \begin{change}12\end{change} different resources, including the Media player, the Camera, and the Location manager.

For each case, we used Appium~\cite{Appium_2017}, a test automation tool for mobile apps, to manually implement an automatic test case that reproduces the sequence of actions that has been reported to produce the misuse. All the executions have been performed on an actual smartphone: \begin{change}the Samsung Galaxy S5 smartphone running the %\footnote{\begin{change}We used  CyanogenMod to be able to run Android 6.0.1 in our device. Note that this decision has no effect on the lifecycle of the apps and the usage of the API.\end{change}}
 Android 6.0.1 Marshmallow mobile platform, equipped with a 2.5 GHz quad-core Snapdragon 801 system-on-chip, 2 GB of RAM, and 16 GB of internal storage. \end{change}
%Examined applications are affected by at least one misuse for a total of 5 misuses and we executed, for each application, a test case that reproduces such incorrect uses of API. The policies involved in the evaluation are defined in~\cite{Wu:Callback:TSE:2016} and eight of these are triggered by the execution of tests. 

%The mobile device used in this evaluation is a Samsung's Galaxy Nexus smartphone running CyanogenMod 13.0.1 OS based on the Android 6.0.1 Marshmallow mobile platform. This smartphone is equipped with a 1.2 GHz dual-core ARM Cortex-A9 processor, 1 GB of RAM and 16 GB of internal storage.

\begin{table*}[]
\caption{Effectiveness of Proactive Libraries} %\vspace{-0.5cm}
\label{tab:effectiveness}
\resizebox*{0.96\textwidth}{!}{%
\centering

%\begin{tabular}{l|c|l|c|c|}
\begin{tabular}{ccccc}
\toprule
%\multicolumn{1}{c}{\textbf{APP}}                        & {  \textbf{Android Component}}            & \multicolumn{1}{c}{\textbf{API}}                                 & \textbf{Usage Policy}                                                                                                                               & \textbf{\begin{tabular}[c]{@{}c@{}}Policy\\ Violations\end{tabular}}          \\ \midrule

\multicolumn{1}{c}{\textbf{APP}}                        & {  \textbf{Lifecycle Object}}            & \multicolumn{1}{c}{\textbf{API}}                                 & \textbf{Usage Policy}                                                                                                                               & \textbf{Policy Violation}          \\ \midrule

{{  AndroidHacks}} & { Activity} & {  android.app.Activity}                       & {  \begin{tabular}[c]{@{}c@{}}If managedQuery() is invoked, replace \\ it with query() of ContentResolver\end{tabular}}                   & {  \textbf{healed}}                                        \\ \midrule

{BlueChat}                            & {  Activity}                              & android.bluetooth.BluetoothAdapter                                & \begin{tabular}[c]{@{}c@{}}If enable() is invoked, invoke \\ disable() when onDestroy()\end{tabular}                                                         & \textbf{healed}                                                               \\ \midrule

{CheDengWo}                           & {  Activity}                              & android.media.MediaPlayer                                         & \begin{tabular}[c]{@{}c@{}}If \textless{}init\textgreater is invoked, invoke\\ release() when onPause\end{tabular}                                    & \textbf{healed}                                                               \\ \midrule

{ErWeiMaL}                            & {  Activity}                              & android.media.MediaPlayer                                         & \begin{tabular}[c]{@{}c@{}}If \textless{}init\textgreater\ is invoked, invoke \\ release() when onPause\end{tabular}                                   & \textbf{healed}                                                               \\ \midrule
{FontMaster}                          & {  Activity}                              & android.util.LruCache                                             & \begin{tabular}[c]{@{}c@{}}If \textless{}init\textgreater\ is invoked, invoke\\  evictAll() when onDestroy()\end{tabular}                                     & \textbf{healed}                                                               \\ \midrule

%%%% Fiicam
                                   & {  Activity}                              & android.hardware.Camera                                           & \begin{tabular}[c]{@{}c@{}}If open() is invoked, invoke\\ release() when onPause()\end{tabular}                                                              & \begin{tabular}[c]{@{}c@{}}no\\ violation\end{tabular}                        \\ \cmidrule{2-5} 
{\multirow{-4}{*}{Foocam}}            & {  Activity}                              & android.hardware.Camera                                           & \begin{tabular}[c]{@{}c@{}}If startPreview() is invoked, invoke\\ stopPreview() when onPause()\end{tabular}                                                  & \textbf{healed}                                                               \\ \midrule

%%%% GetBackGPS

                    & {  Activity}                              & android.media.MediaPlayer                                         & \begin{tabular}[c]{@{}c@{}}If create() is invoked, invoke \\ release() when onPause\end{tabular}                                                             & \textbf{healed}                                                               \\ \cmidrule{2-5} 
{\multirow{-3}{*}{FromCat}}           & {  MediaRecorder} & \begin{change}android.hardware.Camera\end{change}                                           & \begin{tabular}[c]{@{}c@{}}\begin{change}If MediaRecorder.start() is invoked,\end{change}\\ \begin{change}do not invoke lock()\end{change}\end{tabular}                                                          & \begin{tabular}[c]{@{}c@{}}\begin{change}no\end{change}\\ \begin{change}violation\end{change}\end{tabular}                        \\ \midrule

	& {  Activity}                              & android.hardware.SensorManager                                    & \begin{tabular}[c]{@{}c@{}}If registerListener() is invoked, invoke \\ unregisterListener() when onPause()\end{tabular}                                      & \begin{tabular}[c]{@{}c@{}}no\\ violation\end{tabular}                        \\ \cmidrule{2-5} 
 
	& {  Activity}                              & android.location.LocationManager                                  & \begin{tabular}[c]{@{}c@{}}If requestLocationUpdates() is invoked, \\ invoke removeUpdates() when onPause()\end{tabular}                          & \begin{tabular}[c]{@{}c@{}}no\\ violation\end{tabular}                        \\ \cmidrule{2-5} 
	
	& {  Activity}                              & android.os.RemoteCallbackList                                     & \begin{tabular}[c]{@{}c@{}}If register() is invoked, invoke\\ unregister() when onPause()\end{tabular}                                                       & \begin{tabular}[c]{@{}c@{}}no\\ violation\end{tabular}                        \\ \cmidrule{2-5} 
	& {  Service}                               & {  android.hardware.SensorManager}             & {  \begin{tabular}[c]{@{}c@{}}If registerListener() is invoked, invoke\\ unregisterListener() when onDestroy()\end{tabular}}             & {  \begin{tabular}[c]{@{}c@{}}no\\ violation\end{tabular}} \\ \cmidrule{2-5} 
	
	& {  Service}                               & {  android.location.LocationManager}           & {  \begin{tabular}[c]{@{}c@{}}If requestLocationUpdates() is invoked, \\ invoke removeUpdates() when onDestroy()\end{tabular}} & {  \begin{tabular}[c]{@{}c@{}}no\\ violation\end{tabular}} \\ \cmidrule{2-5} 
{\multirow{-12}{*}{GetBackGPS}}                                   & {  Service}                               & {  android.os.RemoteCallbacklist}              & {  \begin{tabular}[c]{@{}c@{}}If register() is invoked, invoke \\ unregister() when onDestroy()\end{tabular}}             & {  \begin{tabular}[c]{@{}c@{}}no\\ violation\end{tabular}} \\ \midrule

%%%%% IPST2
{  IPST2}        & {  Resources} & {  android.content.res.Resources}              & {  \begin{tabular}[c]{@{}c@{}}If getDrawable is invoked, replace it with\\ getDrawable of AppCompatDrawableManager\end{tabular}}      & {  \textbf{healed}}                                        \\ \midrule

%%%%% Mama
{MaMa}                                & {  Activity}                              & android.media.MediaPlayer                                         & \begin{tabular}[c]{@{}c@{}}If \textless{}init\textgreater\ is invoked, invoke \\ release() when onPause\end{tabular}                                          & \textbf{healed}                                                               \\ \midrule

%%%%% QiCaiScan
	& {  Activity}                              & android.hardware.Camera                                           & \begin{tabular}[c]{@{}c@{}}If open() is invoked, invoke \\ release() when onPause()\end{tabular}                                                             & \textbf{healed}                                                               \\ \cmidrule{2-5} 
	& {  Activity}                              & android.hardware.Camera                                           & \begin{tabular}[c]{@{}c@{}}If startPreview() is invoked, invoke\\ release() when onPause()\end{tabular}                                                      & \textbf{healed}                                                               \\ \cmidrule{2-5} 
	{\multirow{-6}{*}{QiCaiScan}}         & {  Activity}                              & android.media.MediaPlayer                                         & \begin{tabular}[c]{@{}c@{}}If \textless{}init\textgreater\ is invoked, invoke \\ release() when onPause\end{tabular}                                          &  {  \begin{tabular}[c]{@{}c@{}}no\\ violation\end{tabular}} \\ \midrule

	%\begin{change2b}no violation\end{change2b}                                                               \\ \midrule

%%%%% SuperTorch
	& {  Activity}                              & android.media.MediaPlayer                                         & \begin{tabular}[c]{@{}c@{}}If \textless{}init\textgreater\ is invoked, invoke \\ release() when onPause\end{tabular}                                          & \textbf{healed}                                                               \\ \cmidrule{2-5} 
{\multirow{-3}{*}{SuperTorch}}        & {  Activity}                              & android.location.LocationManager                                  & \begin{tabular}[c]{@{}c@{}}If requestLocationUpdates() is invoked,\\ invoke removeUpdates() when onPause()\end{tabular}                            & \textbf{healed}                                                               \\ \midrule
	 & {  Activity}                              & android.media.MediaPlayer                                         & \begin{tabular}[c]{@{}c@{}}If \textless{}init\textgreater\ is invoked, invoke\\ release() when onPause\end{tabular}                                           & \textbf{healed}                                                               \\ \cmidrule{2-5}

{\multirow{-3}{*}{WebPCSuite}}        & {  Activity}                              & android.os.PowerManager.WakeLock                                  & \begin{tabular}[c]{@{}c@{}}If acquire() is invoked, invoke \\ release() when onPause()\end{tabular}                                                          & \textbf{healed}                                                               \\ \midrule
	
{{  WiFiSaver}}    & {  Service}                               & {  java.lang.Thread}                           & {  \begin{tabular}[c]{@{}c@{}}If \textless{}init\textgreater\ is invoked, invoke\\ interrupt() when onDestroy\end{tabular}}                & {  \textbf{healed}}                                        \\ \midrule
	& {  Activity}                              & android.net.wifi.WifiManager.MulticastLock                        & \begin{tabular}[c]{@{}c@{}}If acquire() is invoked, invoke\\ release() when onDestroy()\end{tabular}                                                         & \begin{tabular}[c]{@{}c@{}}no\\ violation\end{tabular}                        \\ \cmidrule{2-5}

{\multirow{-4}{*}{XiaoMiWiFi}}        & {  Service}                               & {  android.net.wifi.WifiManager.MulticastLock} & {  \begin{tabular}[c]{@{}c@{}}If acquire() is invoked, invoke\\ release() when onDestroy()\end{tabular}}                                  & {  \textbf{healed}}                                        \\ \bottomrule
\end{tabular}
}
\end{table*}

\subsection*{RQ1 - Effectiveness}
To evaluate the effectiveness of proactive libraries, we first identified the usage policies relevant to the APIs involved in the reported misuses and we then defined the corresponding enforcement strategies. To identify the policies, we exploited the information about the use of the Android API reported in~\cite{Liu:ResourceLeaks:ISSRE:2016,Wu:Callback:TSE:2016,AndroidAPI_2017}. Each enforcement strategy has been codified as an edit automaton using the EMEditor and finally turned into a Java proactive module with the PMGenerator. The code generated for the proactive modules ranged from 1K lines of code in the case of \texttt{android.os.PowerManager.WakeLock} to 2K lines of code in the case of \texttt{android.media.MediaPlayer}. Quantitatively most of the code has been generated by the platform dependent rules.

To study the effectiveness of the proactive modules, we executed the test cases that should produce the misuses, and checked whether the misuses have been automatically detected and healed by the proactive modules. Table~\ref{tab:effectiveness} reports the results. Column \emph{APP} indicates the name of the Android app used in the evaluation. \begin{change}Columns \emph{Lifecycle Object} and\end{change} \emph{API} \begin{change}indicate the Android component that has been reported to misuse the API and\end{change} the API that has been reported to be misused\begin{change}, respectively\end{change}. Note that each API maps to a different local resource. 

Column \emph{Usage Policy} indicates the policy that has been reported to be violated by the app. We have written the policies in the form\begin{change}s\end{change} ``if \texttt{methodA} is invoked, invoke \texttt{methodB} when \texttt{callback}'', \begin{change}``if \texttt{methodA} is invoked, replace it with \texttt{methodB}'', and ``if \texttt{methodA} is invoked, do not invoke \texttt{methodB}''\end{change}. The \begin{change}first\end{change} policy should be interpreted as: if the app invokes \texttt{methodA}, it should also invoke \texttt{methodB} when \texttt{callback} is produced by the Android framework, unless \texttt{methodB} has been already invoked before. This policy requires interrupting any ongoing usage of a resource if certain events occur, such as the suspension or destruction of the current activity. \begin{change}The second policy should be interpreted as: if the app invokes \texttt{methodA}, the call should be replaced with a call to \texttt{methodB}. This policy requires replacing calls to faulty or deprecated methods with calls to methods that properly implement the required logic. The third policy should be interpreted as: if the app invokes \texttt{methodB} after  \texttt{methodA}, suppress the call to \texttt{methodB}. This policy prevents the violation of API usage protocols.\end{change}

%%%%%%%%%%%%%%%%%%%%%%%%%%%%%%%%%%%%%%%%%%%%%%%%%%%%%%%%To Review

%Column \emph{Size of EFSM alphabet}  indicates the size of the alphabet of the edit automaton, which corresponds to the number of methods that are intercepted by the proactive module. Note that the set of methods changes based on the API and based on the specific enforcement strategy. In fact, although inserting a method invocation that has not been timely produced by an app (e.g., invoking \texttt{stopPreview()} when a destroyed activity omits to interrupt the generation of the preview) is almost the same for all the enforcement models, the way a resource that has been forcedly released can be transparently reassigned to an app when its execution is resumed changes based on the specific resource type. This aspect produces significantly different enforcement models.

Column \emph{Policy Violations} indicates the effect of the proactive modules on the execution: \emph{healed} indicates that the execution has been automatically corrected, while \emph{no violation} indicates that the policy has not been violated by the execution. In none of the cases the proactive modules fail to heal an execution that violates a resource usage policy. This result confirms the effectiveness and suitability of proactive modules to enforce correctness policies.

It is worth discussing the \begin{change2b}ten\end{change2b} cases where no violation has been detected. In the case of the \emph{fooCam} app, the camera release policy has been reported as violated in~\cite{Liu:ResourceLeaks:ISSRE:2016}, where a static analysis tool is exploited to detect erroneous accesses to resources. We discovered that this case in reality is a false positive produced by static analysis. The path that may violate the resource usage policy is infeasible, thus it can be never executed. This case is correctly captured by the enforcement model that does not report any violation when the app is executed, \begin{change}contrary\end{change} to static analysis. \begin{change2b}Similarly, \emph{FromCat} and \emph{QiCaiScan} apps happened to use the camera properly.\end{change2b}

In the cases belonging to \emph{GetBackGPS} and \emph{XiaoMiWiFi}, which were also reported as violating the corresponding resource usage policies in~\cite{Liu:ResourceLeaks:ISSRE:2016}, we found out that they are not actual policy violations. %Our proactive libraries \begin{change}check the identity of the component that interacts with the library\end{change} before enforcing a policy. 
%hether a resource is acquired within an activity before enforcing a policy since policies define the use of resources in the context of the lifecycle of an activity. 
\begin{change}In these four cases Android activities do not violate any policy because\end{change} the resources are in reality acquired and released by Android services, which are usually used to perform long-term operations in background. %Thus, since the resource is under the control of the service and is not under the control of the Activity there is no violated policy. 
Enforcing the release of the resources in these situations would produce misbehaviours since resources would be released while they are still in use by services. Our proactive modules for activities correctly captured this situation. \begin{change}To confirm this result we also generated the proactive modules checking that Android services acquire and release the resources properly and no violation has been reported.\end{change}

\begin{figure*}%[ht!]
\begin{center}
  \includegraphics[width=12cm]{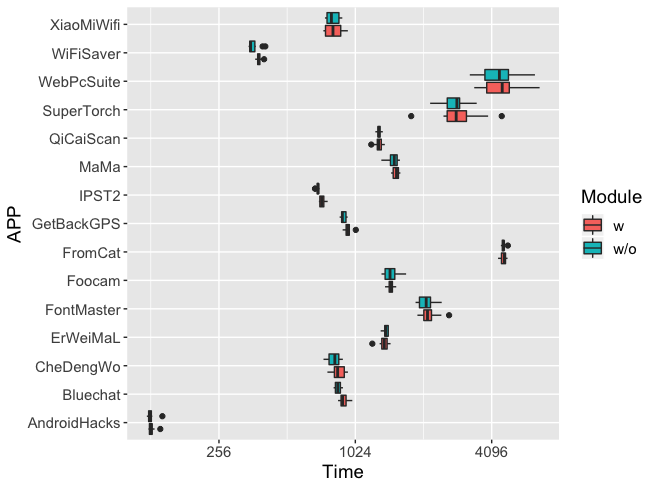}
\caption{Distribution of execution time w and w/o proactive modules }
\label{fig:overhead}
\end{center}
\end{figure*}

%Although the two solutions are designed for completely different scenarios, improving the in-house verification process one and making the end-user environment more dependable regardless the quality of the executed applications the other, they show the typical complementarity of techniques based on static and dynamic analysis, with the former suffering from false positives and the latter suffering from incompleteness.

\subsection*{RQ2 - Overhead}
To measure overhead, we measured the time spent to run each test \begin{change}case\end{change}, when proactive modules are active and when they are not. In the case of apps with multiple policies to be enforced, the experiment was conducted by activating all the proactive modules applicable to the app. For example, in the case of \emph{Foocam}, there are  two modules that act at the same time when the activity is paused: one to enforce the release of the camera, and the other to enforce stop capturing and drawing preview frames.  

\begin{change}In addition to study the scalability of the approach, we measured how the overhead increases for an increasing number of monitored events. For this task, we selected the app whose proactive modules introduced the highest overhead, that is \emph{Bluechat}, and we deployed an increasing number of enforcers. We measured the overhead for every configuration and report the results.\end{change}

Figure~\ref{fig:overhead} \begin{change}shows the overhead measured when the proactive modules that enforce the policies listed in Table~\ref{tab:effectiveness} are active\end{change} with a box plot. Each box represents a population of 10 samples obtained by running the same test case 10 times. We put next to each other the two sets of executions that we collected: the box labeled with \texttt{w/o} shows the runtime cost for the samples collected without activating the proactive module, while the box labeled with \texttt{} shows the runtime cost for the samples collected when the proactive module is active. 

We can notice that in most cases the use of proactive libraries does not generate any measurable overhead. \begin{change}This is confirmed with an ANOVA test that only reports the app as a significant factor (p-value = $2e^{-16}$) and reports the presence of the proactive module as largely non-significant (p-value = $0.99$). Overhead is low\end{change} because the enforcement models for the majority of the cases leave the input unaltered or perform very simple operations. Considering all \begin{change}15 apps\end{change} we have an average overhead of 2\%, \begin{change}even if in several cases (7 out of 15 apps) we have multiple modules active at the same time: 2 proactive modules for \emph{Foocam}, \emph{FromCat}, \emph{SuperTorch}, \emph{WebPCSuite}, and \emph{XiaoMiWiFi}, 3 proactive modules for \emph{QiCaiScan}, and 6 proactive modules for \emph{GetBackGPS}. \end{change}

%In 5 cases the observed overhead is the same than the time that would be required by the app to release the resource correctly, and 
In the worst case we reported an overhead equals to \begin{change}6\% (the actual overhead was 52$ms$ in the context of a functionality that required 858$ms$ to complete)\end{change}, which is still an amount of overhead that can be hardly perceived~\cite{NIER17}. \begin{change}The fact that the overhead introduced with the proactive modules does not introduce significant differences in the execution time of the apps has been confirmed with an ANOVA test that revealed no significant differences (with a significance level equals to $0.05$) between the app with and without the enforcers running. \end{change}

\begin{figure*}%[ht!]
\begin{center}
  \includegraphics[width=12cm]{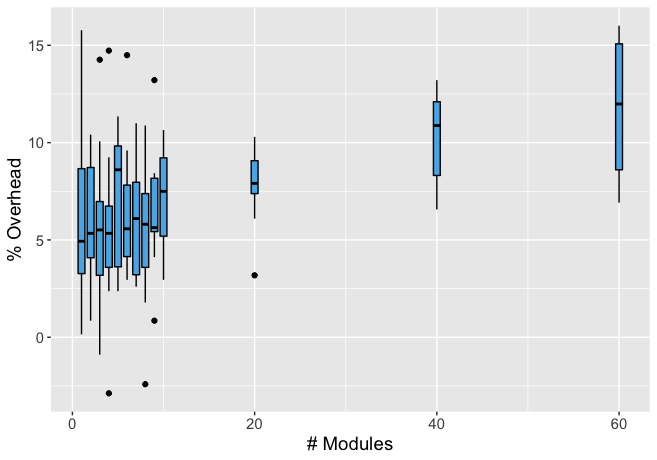}
\caption{Overhead vs number of proactive modules}
\label{fig:scalability}
\end{center}
\end{figure*}

\begin{change}Figure~\ref{fig:scalability} shows the overhead (y axis) observed for an increasing number of proactive modules deployed on the app (x axis).  Each box is the result of 10 executions. It is possible to notice how the approach scales well with the deployment of multiple proactive modules and thus the monitoring of many events. The presence of the monitoring infrastructure causes a mean overhead around 5\% already when a single proactive module is deployed. The overhead however growths smoothly with the number of active modules. Indeed it is only slightly higher when 60 enforcers have been deployed reaching a mean value around 12\%. This result shows that the approach does not suffer any particular limitation in terms of the number of enforcement strategies that can be simultaneously active.\end{change}

\medskip

Based on these results, we can conclude that enforcers can be feasibly used by final users to improve the reliability of their environment. %, despite the presence of untrusted software apps. %Although we did not try with a large number of policies simultaneously active, the cost of monitoring these properties is likely to be affordable. In principle, only the actions that produce the simultaneous violation of many policies might be affected by a significant overhead due to the simultaneous insertion of several actions necessary to enforce all the violated policies at once, but again the total execution time would be similar to the execution time required by a correct implementation of the app. 

\begin{change}
\subsection*{RQ3 - Usefulness}
To investigate if the modeling and code generation environment is actually useful compared to implementing Xposed modules from scratch we defined an experiment involving human subjects. In particular, we involved three developers \begin{change2}who graduated from our department\end{change2} with different levels of \begin{change2}experience\end{change2} with Android and Xposed: an \emph{\begin{change2}Experienced\end{change2} User} with more than 1 year of experience with both Android and Xposed, a \emph{\begin{change2}Moderately Experienced\end{change2} User} with some months of expertise with both Android and Xposed, finally a \emph{Novice to Xposed} user with years of expertise of Android development but only a 15 minutes training on Xposed. 

After a 15mins debriefing on edit automata and our modeling environment, we asked each subject \begin{change2}for each app\end{change2} to first design the enforcement strategy in our MDSD environment and then implement the same strategy with Xposed. We fixed a time limit of 3 hours for each task. Note that this setting is very favourable to the implementation of the modules directly with Xposed because the design effort necessary to identify the right strategy is entirely spent in the modeling phase and when using Xposed the subjects already know the enforcement strategy they have to implement. Each subject worked on two apps and two policies, one policy that refers to Android activities and one policy that refers to Android services. \begin{change2}To prevent human subjects from being tired while performing their tasks, the four tasks have been performed over four days, one task per day.\end{change2} We measure the time required to complete the task in the two cases for each subject. \begin{change2}Finally, note that the human subjects had to use the actual apps only to validate their implementation, which can be anyway designed and implemented (generated in the case of our MDSD environment) regardless of the target app. In fact, the enforcement strategies require no information about the target apps that may violate the policies.\end{change2}

Table~\ref{tab:effort} reports the results. \begin{change2}When using our MDSD environment\end{change2}, all the subjects managed to obtain correct enforcers in less than 1 hour, with the \begin{change2}experienced\end{change2} user completing the task faster than all the other subjects. On the contrary only the \begin{change2}experienced\end{change2} user managed to implement correct enforcers in less than three hours \begin{change2}when using Xposed\end{change2}. This is mainly due to the complexity of the interactions between the monitored components, which is well abstracted in our modeling environment. In fact, the implementation has to deal with aspects such as tracing the identity of the components that interact with the considered APIs, transparently resuming resources that have been released, and avoiding the interference between the activity of the enforcer and the Xposed module. We can finally notice that although the \begin{change2}experienced\end{change2} user finally obtained the module programmatically, he needed an order of magnitude more time than the time needed using our MDSD environment to achieve it. These results suggest that our MDSD environment provides a proper abstraction of the problem of designing an enforcement strategy and it can be useful to \begin{change2}experienced\end{change2} users, to quickly obtain enforcers that can be deployed, and \begin{change2}to both moderately experienced and\end{change2} novice users, who can more easily approach the technology.
\end{change}

%
%\begin{table*}[]
%\begin{change}
%\small
%\centering
%\scriptsize
%\caption{Effort comparison}
%\label{tab:effort}
%\begin{tabular}{|l| l| c| c|| c| c|}
%\hline
%
%\textbf{Subject} & \textbf{App} & \textbf{Android Component} & \textbf{Policy} & \textbf{Time with MDSD} & \textbf{Time with Xposed} \\
%\hline
%
%\multirow{3}{*}{\emph{Expert User}}   & SuperTorch & Activity & \begin{tabular}[c]{@{}c@{}}If \textless{}init\textgreater is invoked, invoke\\ release() when onPause\end{tabular} &  15 min &  2 h 51 min  \\
%						& WiFiSaver & Service &  \begin{tabular}[c]{@{}c@{}}If \textless{}init\textgreater is invoked, invoke\\ interrupt() when onDestroy\end{tabular} &  16 min & 2 h 25 min  \\
%\hline
%
%\multirow{3}{*}{\emph{Skilled User}}   & SuperTorch & Activity & \begin{tabular}[c]{@{}c@{}}If \textless{}init\textgreater is invoked, invoke\\ release() when onPause\end{tabular} &  32 min & $>$3h  \\
%						& WiFiSaver & Service &  \begin{tabular}[c]{@{}c@{}}If \textless{}init\textgreater is invoked, invoke\\ interrupt() when onDestroy\end{tabular} &  25 min & $>$3h   \\
%\hline
%
%\multirow{3}{*}{\emph{Novice to XPosed}}   & SuperTorch & Activity & \begin{tabular}[c]{@{}c@{}}If \textless{}init\textgreater is invoked, invoke\\ release() when onPause\end{tabular} &  45 min & $>$3h  \\
%						& WiFiSaver & Service &  \begin{tabular}[c]{@{}c@{}}If \textless{}init\textgreater is invoked, invoke\\ interrupt() when onDestroy\end{tabular} &  15 min & $>$3h  \\
%\hline
%
%\end{tabular}
%\end{change}
%\end{table*}
%

\begin{table*}[]
\begin{change}
\caption{Effort comparison}
\label{tab:effort}
%\resizebox{\textwidth}{!}{%
\resizebox{15cm}{!}{%
  \centering
\begin{tabular}{cccc|cc}
        \toprule

%\begin{table*}[]
%\begin{change}
%\small
%\centering
%\scriptsize
%\caption{Effort comparison}
%\label{tab:effort}
%\begin{tabular}{|l| l| c| c|| c| c|}
%\hline

\textbf{Subject} & \textbf{App} & \textbf{Lifecycle Object} & \textbf{Policy} & \textbf{Time with MDSD} & \textbf{Time with Xposed} \\
\midrule

\multirow{3}{*}{\emph{\begin{change2}Experienced\end{change2} User}}   & SuperTorch & Activity & \begin{tabular}[c]{@{}c@{}}If \textless{}init\textgreater is invoked, invoke\\ release() when onPause\end{tabular} &  15 min &  2 h 51 min  \\ \cmidrule{2-6}
						& WiFiSaver & Service &  \begin{tabular}[c]{@{}c@{}}If \textless{}init\textgreater is invoked, invoke\\ interrupt() when onDestroy\end{tabular} &  16 min & 2 h 25 min  \\
\midrule

\multirow{3}{*}{\emph{\begin{change2}Moderately Experienced\end{change2} User}}   & SuperTorch & Activity & \begin{tabular}[c]{@{}c@{}}If \textless{}init\textgreater is invoked, invoke\\ release() when onPause\end{tabular} &  32 min & $>$3h  \\ \cmidrule{2-6}
						& WiFiSaver & Service &  \begin{tabular}[c]{@{}c@{}}If \textless{}init\textgreater is invoked, invoke\\ interrupt() when onDestroy\end{tabular} &  25 min & $>$3h   \\
\midrule

\multirow{3}{*}{\emph{Novice to XPosed}}   & SuperTorch & Activity & \begin{tabular}[c]{@{}c@{}}If \textless{}init\textgreater is invoked, invoke\\ release() when onPause\end{tabular} &  45 min & $>$3h  \\ \cmidrule{2-6}
						& WiFiSaver & Service &  \begin{tabular}[c]{@{}c@{}}If \textless{}init\textgreater is invoked, invoke\\ interrupt() when onDestroy\end{tabular} &  15 min & $>$3h  \\
\bottomrule
\end{tabular}
}
\end{change}
\end{table*}

\begin{change}
\subsection*{Threats to Validity}
The main internal threats to validity of our experiments relate to the definition of the enforcement model. In RQ1 and RQ2 we studied the applicability of proactive libraries using enforcement models designed by us. Although the enforcement strategies that we used are not particularly sophisticated at the model level, different users may design different strategies. To mitigate this issue we investigated the design of enforcement strategies with RQ3. All the subjects produced the same enforcement strategy consistently with the one designed by us when using our MDSD tool. In the case of the overhead, we further mitigated the issue by studying the scalability of the approach with respect to a number of configurations.     

The investigation of RQ3 required the involvement of developers. To mitigate the issue about the background knowledge of the developers, we controlled this factor explicitly. To derive a worst case result, we also designed our study such a way that the developers implementing the Xposed modules already know how the enforcer should behave based on their experience with the MDSD tool. In reality the gap between subjects who use our modelling environment and subjects who implement modules directly would be higher.

The main external threat to validity is about the generalizability of the results. We considered various apps, faults, and strategies to mitigate this issue. The results that we achieved are quite consistent across cases so far, thus we expect a good degree of generalization, while we cannot exclude the existence of cases that can challenge our approach, as discussed in the next section. 
\end{change}

\begin{change}
\subsection*{Discussion}
Proactive libraries can be a useful technology to deal with library misuses. In fact, if final users notice any misbehaving app, they can activate the available proactive modules to increase the reliability of their apps at the cost of introducing little overhead, as shown in our experiments. \begin{change2}Note that these bugs may require a significant amount of time to be fixed. Out of the three bugs affecting open source apps, two of these bugs are still open at the time the paper was submitted and one bug has been fixed after 2 years and 3 months.\end{change2}

\begin{change2}The evaluation confirmed the generality of proactive modules. In fact, enforcement strategies are defined using app-independent information only: information about the life-cycle of Android components and information about the library APIs. As a consequence, a same policy is always addressed with a same proactive module, which is reused across multiple apps. This is exactly what happened in our evaluation where we addressed the 27 cases reported in Table~\ref{tab:effectiveness} with only 19 proactive modules, always reusing a same module across different apps when a same policy has been addressed. \end{change2} 

Indeed proactive modules cannot be used to address any possible problem present in mobile applications. A first limitation is the direct consequence of their quality of being app independent. Since a proactive module relies uniquely on the library API and the app lifecycle events it can be applied to any Android app. On the other hand, app independence has a cost, that is, proactive modules cannot be used to deal with problems that require the knowledge of the app to be fixed, such as faults that do not imply any interaction with libraries but only depend on the internal behavior of the app.

Proactive libraries, as demonstrated in the empirical evaluation, can deal with multiple Android components, such as activities and services. The approach is not conceptually limited to these two components, but it can deal with any component and class interacting with a library. A limitation of the current implementation of our MDSD tool is that a same enforcement strategy cannot include multiple clients for a same library API. This limitation can be eliminated by extending the language of the edit automata and the corresponding tool support.

Another source of limitation is the need of intercepting events at run-time to inject and suppress events. In our prototype implementation we used Xposed, which requires rooting the device. This solution is not always practical because users might be unhappy with this requirement. However, there are also other options to affect the run-time behavior of apps. For example, apps can be instrumented in the smartphone, so that they can be monitored and enforced without requiring rooting the device, as done in the work by Neisse et al.~\cite{Neisse:EnforcingPrivacyAndroid:ComputerSecurity:2016}.   We are considering this alternative implementation also for proactive modules.   

Some events might be particularly hard to detect. For instance, peculiar app implementations such as the case of apps that bypass a library API producing native calls would not be intercepted by a proactive module. In general, the events that can be handled by proactive modules are either Java API calls or lifecycle callback events.      
\end{change}

\section{Related Work}\label{sec:related} 
Android is one of the most popular software ecosystems that allows developers to easily develop and distribute apps, and allows users to easily retrieve and install applications. However, the development of reliable apps is a complex task that has to face several challenges, such as the unpredictability and the high variability of the environment~\cite{Muccini_STM_2012, Amalfitano_CCE_2013}, the instability and rapid evolution of the software platform \cite{McDonnell_ESA_2013, LinaresVasquez_ACF_2013}, as well as the lack of strong and reliable analyses and testing tools \cite{Joorabchi_RCM_2013, Khalid_PDT_2014}. This often results in poorly developed applications, which may cause other apps or even the entire system to perform badly and crash.

In this paper we presented an approach that allows users to protect their environment from unreliable apps that improperly use APIs by enforcing the correct usage protocols. There are three distinct areas of research that are related to our contribution: detection of API misuses, incorrect resource management, and design of self-healing systems. 

\smallskip

\emph{API misuses}. Failures caused by API misuses are very popular~\cite{Amani:MUBench:MSR:2016}. These failures have been proven to be pervasive in software systems and different automated approaches are available to detect different types of misuses~\cite{Mariani:BCT:TSE:2011,Wasylkowski:MiningTemporalSpec:ASE:2009,Li:PRMiner:ESECFSE:2005}. However, it is still impossible to completely prevent API misuses, due to limitations of the available techniques that can only detect specific classes of faults.

The presence of applications that incorrectly interact with an API could be further amplified by the rapid evolution of APIs, in fact developers often struggle to adapt their software to the latest versions of libraries. These problems are well-known to affect modern ecosystems, such as Android~\cite{Linares-Vasquez:AndoirdAPIChange:ESECFSE:2013,Egele:CrytographicAndroidMisuse:CCS:2013}. The likelihood of encountering any of these problems while using an app downloaded from a marketplace is further increased by the lack of control over the process used to develop the apps available on the market. In fact, contributors are often more focused on rapidly prototyping their apps than developing high-quality applications. In a nutshell, despite the existence of techniques to assess interactions with APIs, problems due to API misuse are still often encountered by users~\cite{Azim_Towards_2014,Banerjee:EnergyBugs:FSE:2014,Wei:AndroidFragmentation:ASE:2016,Shan:ResumerRestartErrors:OOPSLA:2016,Wu:Callback:TSE:2016}.

This paper proposes to increase the reliability of devices that use applications developed by potentially %untrusted
third parties replacing traditional reactive libraries with proactive libraries, which can still be executed as reactive libraries when needed. Proactive libraries can transform the interaction between an API and unsafe applications to ensure that they do not violate usage protocols. This strategy can be effective in preventing multiple classes of problems, especially those related to the use of resources. 

%do not make any assumption about the way apps have been developed and can be used to fix common problems, especially the ones that may cause issues across apps, such as accessing resources according to wrong patterns.

The proactive libraries share the idea of extending libraries with ReBa~\cite{Dig:ReBa:ICSE:2008}, which is a technique for the development of libraries augmented with adapters to ensure backward compatibility. However, ReBa and the proactive libraries have different goals and adopt different solutions. ReBa addresses the problems introduced by software evolution, while proactive libraries are a general mechanism for enforcing policies at runtime. In addition, ReBa is purely reactive, while proactive libraries use proactive modules to timely take actions to ensure correctness. Compared to a purely reactive solution, proactive modules can change executions more broadly by accessing information that otherwise cannot be intercepted. 

\smallskip

\emph{Bad resource management}. Although the concept of proactive library is general, \begin{change}many of the cases studied in the paper are about\end{change} failures caused by bad resource management, such as apps that acquire and release resources according to wrong patterns~\cite{Riganelli:SPE:ResourceLeak:2019,Azim_Towards_2014,Li:GreenProgramming:GREENS:2014,Wu:Callback:TSE:2016,Liu:ResourceLeaks:ISSRE:2016}. Although some of these problems may be discovered with ad-hoc tests and static analysis techniques~\cite{Azim_Towards_2014,Wu:Callback:TSE:2016,Liu:ResourceLeaks:ISSRE:2016}, it is generally difficult to eliminate resource management problems, covering every possible situation, also considering the problem of fragmentation that affects Android~\cite{Wei:AndroidFragmentation:ASE:2016}. In addition, installing apps that misuse resources can also impact apps that properly interact with the same resources, increasing the need for solutions that can operate in the device. Our empirical results suggest that the reliability of interaction with resources can significantly improve if proactive libraries are adopted to replace reactive libraries, \begin{change}thus providing to library developers and app users a solution that complements the in-house testing and verification activities\end{change}. 

\smallskip

\emph{Self-healing}. Techniques for avoiding and mitigating the impact of failures have been studied in many different contexts, including Web applications~\cite{Magalhaes_SSH_2015}, cyber-physical systems~\cite{Seiger2016}, and Cloud computing~\cite{Dai_SHD_2009}. Some techniques have also explored the idea of detecting and reacting to wrong situations, for example by using exception handlers~\cite{Chang:HealingConnectors:TOSEM:2013,Chang:HealingConnectors:ICSE:2009} and adapters~\cite{Denaro:TestAndAdapt:TOSEM:2013}. However, only a few self-healing solutions have been designed to address an environment with limited resources such as a mobile device.

The first results in the Android domain were mainly related to dynamic patch injection, automatic suppression of faulty functionality, and healing of data loss problems. Dynamic patch injection can be used to quickly deploy fixes in apps~\cite{Mulliner_PST_2013, Zhang_AAG_2014}. However, since patches are produced offline, healing is only achieved with the intervention of the developer, similarly to program repair techniques~\cite{Gazzola:Repair:TSE:2017}. This mechanism can be useful to solve important vulnerabilities, but it cannot be used for immediate and automatic healing of failing executions. 

Automatic suppression mechanisms can be used to automatically detect crashes and avoid future occurrences of the same crashes by bypassing the execution of features that caused the crash.~\cite{Azim_Towards_2014}. This approach may be useful in preventing further failures, but does not help with fixing problems.

Finally, healing of data loss problems provides a strategy to prevent any data loss due to an incorrect implementation of mechanisms to suspend and recover the execution of the apps~\cite{Riganelli:HealingDataLos:IWSF:2016,Riganelli:MSR:DLBenchmark}. 

Compared to these techniques, proactive libraries provide a design solution that is complementary to mechanisms such as dynamic patch injection, and potentially more general than approaches that address specific classes of failures, such as data loss problems.

It is worth mentioning that runtime enforcement in the Android environment has also been investigated by Falcone et al.~\cite{Falcone:Runtime:2012}, but their approach does not take into account the idea of designing policy enforcers that exploit the knowledge of general lifecycle events to increase the effectiveness of the enforcement strategies and does not include automatic code generation.

Finally, there are different approaches for the automatic adaptation of the software running on mobile devices to a changing context~\cite{Elmalaki:CareDroid:MobiCom:2015,Rouvoy:MUSIC:MobMid:2008,Mancinelli:ResourceModel:SEAMS:2006}. Although adaptation mechanisms are usually driven by the need for optimization rather than failure, it might be possible to explore how to apply proactive libraries to enforce adaptation policies rather than correctness policies.

\section{Conclusions}\label{sec:conclusion}
Program libraries implement reusable functionalities that can be conveniently integrated into many different applications. The correctness of this integration depends on the ability to satisfy the assumptions of the library, usually consisting of constraints on when and how certain library operations can be performed. Unfortunately, satisfying these constraints is not always easy. This might be due to the complexity of the APIs implemented by the library, tacit assumptions about the library, and the complexity of the applications that use the library. For example, Android apps are particularly prone to incorrect interactions with their libraries~\cite{Liu:ResourceLeaks:ISSRE:2016,Azim_Towards_2014,Wei:AndroidFragmentation:ASE:2016,Wu:Callback:TSE:2016}.

To address this problem, we introduced the concept of proactive library, which combines a regular reactive library with multiple proactive modules that can monitor executions and enforce correctness policies. These policies define how a library should be properly used at runtime. Since proactive modules are monitors that can alter executions according to a strategy, we propose to model their behavior as enforcement models, which offer a natural formalism to represent how an execution can be modified by suppressing and inserting method calls. 

We use a model-driven software development process to automatically generate the proactive modules from models. In particular, we defined a code generation strategy that separates the code generated directly from the model and the code that depends on the target platform. This allows the code generator to be extended to other target environments while preserving the code generation rules that do not depend on the target platform. 

We evaluated our solution on the Android environment focusing on relevant classes of APIs and correctness policies about access to resources. %The proactive modules obtained from the enforcement models that we defined use Xposed~\cite{Xposed_2016} to intercept events and alter executions. 
Results show that proactive libraries can effectively and efficiently enforce the specified resource usage policies.

\section*{Acknowledgment}
The authors want to thank Jierui Liu, Tianyong Wu, Jun Yan, and Jian Zhang for sharing with us information about the experimental subjects used to evaluate RelFix and for answering our questions.

This work has been partially supported by the H2020 Learn project, which has been funded under the ERC Consolidator Grant 2014 program (ERC Grant Agreement n.646867) and the GAUSS national research project, which has been funded by the MIUR under the PRIN 2015 program (Contract 2015KWREMX).

\bibliographystyle{ACM-Reference-Format}
\bibliography{main} 


\begin{thebibliography}{57}

%%% the Web address.

\ifx \showCODEN    \undefined \def \showCODEN     #1{\unskip}     \fi
\ifx \showDOI      \undefined \def \showDOI       #1{#1}\fi
\ifx \showISBNx    \undefined \def \showISBNx     #1{\unskip}     \fi
\ifx \showISBNxiii \undefined \def \showISBNxiii  #1{\unskip}     \fi
\ifx \showISSN     \undefined \def \showISSN      #1{\unskip}     \fi
\ifx \showLCCN     \undefined \def \showLCCN      #1{\unskip}     \fi
\ifx \shownote     \undefined \def \shownote      #1{#1}          \fi
\ifx \showarticletitle \undefined \def \showarticletitle #1{#1}   \fi
\ifx \showURL      \undefined \def \showURL       {\relax}        \fi
% invisible to TeX
\providecommand\bibfield[2]{#2}
\providecommand\bibinfo[2]{#2}
\providecommand\natexlab[1]{#1}
\providecommand\showeprint[2][]{arXiv:#2}

\bibitem[\protect\citeauthoryear{Amalfitano, Fasolino, Tramontana, and
  Amatucci}{Amalfitano et~al\mbox{.}}{2013}]%
        {Amalfitano_CCE_2013}
\bibfield{author}{\bibinfo{person}{D. Amalfitano}, \bibinfo{person}{A.~R.
  Fasolino}, \bibinfo{person}{P. Tramontana}, {and} \bibinfo{person}{N.
  Amatucci}.} \bibinfo{year}{2013}\natexlab{}.
\newblock \showarticletitle{Considering Context Events in Event-Based Testing
  of Mobile Applications}. In \bibinfo{booktitle}{{\em Proceedings of the
  International Conference on Software Testing, Verification and Validation
  Workshops (ICSTW)}}.
\newblock


\bibitem[\protect\citeauthoryear{Amani, Nadi, Nguyen, Nguyen, and Mezini}{Amani
  et~al\mbox{.}}{2016}]%
        {Amani:MUBench:MSR:2016}
\bibfield{author}{\bibinfo{person}{S. Amani}, \bibinfo{person}{S. Nadi},
  \bibinfo{person}{H.~A. Nguyen}, \bibinfo{person}{T.~N. Nguyen}, {and}
  \bibinfo{person}{M. Mezini}.} \bibinfo{year}{2016}\natexlab{}.
\newblock \showarticletitle{{MUBench}: A Benchmark for API-misuse Detectors}.
  In \bibinfo{booktitle}{{\em Proceedings of the International Conference on
  Mining Software Repositories (MSR)}}.
\newblock


\bibitem[\protect\citeauthoryear{Android}{Android}{2019a}]%
        {Android:Lifecycle:website}
\bibfield{author}{\bibinfo{person}{Android}.} \bibinfo{year}{2019}\natexlab{a}.
\newblock \bibinfo{title}{The Activity Lifecycle}.
\newblock
  \bibinfo{howpublished}{\url{https://developer.android.com/guide/components/activities/activity-lifecycle.html}}.
    (\bibinfo{year}{2019}).
\newblock
\newblock
\shownote{[Online; accessed 4 January 2019].}


\bibitem[\protect\citeauthoryear{Android}{Android}{2019b}]%
        {AndroidAPI_2017}
\bibfield{author}{\bibinfo{person}{Android}.} \bibinfo{year}{2019}\natexlab{b}.
\newblock \bibinfo{title}{{Android API}}.
\newblock
  \bibinfo{howpublished}{\url{https://developer.android.com/guide/index.html}}.
    (\bibinfo{year}{2019}).
\newblock
\newblock
\shownote{[Online; accessed 4 January 2019].}


\bibitem[\protect\citeauthoryear{Android}{Android}{2019c}]%
        {AndroidStudio}
\bibfield{author}{\bibinfo{person}{Android}.} \bibinfo{year}{2019}\natexlab{c}.
\newblock \bibinfo{title}{{Android Studio}}.
\newblock
  \bibinfo{howpublished}{\url{https://developer.android.com/studio/index.html}}.
    (\bibinfo{year}{2019}).
\newblock
\newblock
\shownote{[Online; accessed 21 January 2019].}


\bibitem[\protect\citeauthoryear{Azim, Neamtiu, and Marvel}{Azim
  et~al\mbox{.}}{2014}]%
        {Azim_Towards_2014}
\bibfield{author}{\bibinfo{person}{M.~T. Azim}, \bibinfo{person}{I. Neamtiu},
  {and} \bibinfo{person}{L.~M. Marvel}.} \bibinfo{year}{2014}\natexlab{}.
\newblock \showarticletitle{Towards Self-healing Smartphone Software via
  Automated Patching}. In \bibinfo{booktitle}{{\em Proceedings of the
  International Conference on Automated Software Engineering (ASE)}}.
\newblock


\bibitem[\protect\citeauthoryear{Banerjee, Chong, Chattopadhyay, and
  Roychoudhury}{Banerjee et~al\mbox{.}}{2014}]%
        {Banerjee:EnergyBugs:FSE:2014}
\bibfield{author}{\bibinfo{person}{A. Banerjee}, \bibinfo{person}{L.~Kee
  Chong}, \bibinfo{person}{S. Chattopadhyay}, {and} \bibinfo{person}{A.
  Roychoudhury}.} \bibinfo{year}{2014}\natexlab{}.
\newblock \showarticletitle{Detecting Energy Bugs and Hotspots in Mobile Apps}.
  In \bibinfo{booktitle}{{\em Proceedings of the ACM SIGSOFT International
  Symposium on Foundations of Software Engineering (FSE)}}.
\newblock


\bibitem[\protect\citeauthoryear{Chang, Mariani, and Pezz\`{e}}{Chang
  et~al\mbox{.}}{2009}]%
        {Chang:HealingConnectors:ICSE:2009}
\bibfield{author}{\bibinfo{person}{H. Chang}, \bibinfo{person}{L. Mariani},
  {and} \bibinfo{person}{M. Pezz\`{e}}.} \bibinfo{year}{2009}\natexlab{}.
\newblock \showarticletitle{In-field Healing of Integration Problems with COTS
  Components}. In \bibinfo{booktitle}{{\em Proceedings of the International
  Conference on Software Engineering (ICSE)}}.
\newblock


\bibitem[\protect\citeauthoryear{Chang, Mariani, and Pezz\`{e}}{Chang
  et~al\mbox{.}}{2013}]%
        {Chang:HealingConnectors:TOSEM:2013}
\bibfield{author}{\bibinfo{person}{H. Chang}, \bibinfo{person}{L. Mariani},
  {and} \bibinfo{person}{M. Pezz\`{e}}.} \bibinfo{year}{2013}\natexlab{}.
\newblock \showarticletitle{Exception Handlers for Healing Component-based
  Systems}.
\newblock \bibinfo{journal}{{\em ACM Transactions on Software Engineering and
  Methodologies (TOSEM)\/}} \bibinfo{volume}{22}, \bibinfo{number}{4}
  (\bibinfo{year}{2013}), \bibinfo{pages}{30:1--30:40}.
\newblock


\bibitem[\protect\citeauthoryear{Cornejo, Briola, Micucci, and Mariani}{Cornejo
  et~al\mbox{.}}{2017}]%
        {NIER17}
\bibfield{author}{\bibinfo{person}{O. Cornejo}, \bibinfo{person}{D. Briola},
  \bibinfo{person}{D. Micucci}, {and} \bibinfo{person}{L. Mariani}.}
  \bibinfo{year}{2017}\natexlab{}.
\newblock \showarticletitle{In the field monitoring of software applications}.
  In \bibinfo{booktitle}{{\em Proceedings of the International Conference on
  Software Engineering (ICSE), NIER track}}.
\newblock


\bibitem[\protect\citeauthoryear{Dai, Xiang, and Zhang}{Dai
  et~al\mbox{.}}{2009}]%
        {Dai_SHD_2009}
\bibfield{author}{\bibinfo{person}{Y. Dai}, \bibinfo{person}{Y. Xiang}, {and}
  \bibinfo{person}{G. Zhang}.} \bibinfo{year}{2009}\natexlab{}.
\newblock \showarticletitle{Self-healing and Hybrid Diagnosis in Cloud
  Computing}. In \bibinfo{booktitle}{{\em Proceedings of the International
  Conference on Cloud Computing (CloudCom)}}.
\newblock


\bibitem[\protect\citeauthoryear{Denaro, Pezz\`{e}, and Tosi}{Denaro
  et~al\mbox{.}}{2013}]%
        {Denaro:TestAndAdapt:TOSEM:2013}
\bibfield{author}{\bibinfo{person}{G. Denaro}, \bibinfo{person}{M. Pezz\`{e}},
  {and} \bibinfo{person}{D. Tosi}.} \bibinfo{year}{2013}\natexlab{}.
\newblock \showarticletitle{Test-and-adapt: An Approach for Improving Service
  Interchangeability}.
\newblock \bibinfo{journal}{{\em ACM Transactions on Software Engineering and
  Methodologies (TOSEM)\/}} \bibinfo{volume}{22}, \bibinfo{number}{4}
  (\bibinfo{year}{2013}), \bibinfo{pages}{28:1--28:43}.
\newblock


\bibitem[\protect\citeauthoryear{Dig, Negara, Mohindra, and Johnson}{Dig
  et~al\mbox{.}}{2008}]%
        {Dig:ReBa:ICSE:2008}
\bibfield{author}{\bibinfo{person}{D. Dig}, \bibinfo{person}{S. Negara},
  \bibinfo{person}{V. Mohindra}, {and} \bibinfo{person}{R. Johnson}.}
  \bibinfo{year}{2008}\natexlab{}.
\newblock \showarticletitle{{ReBA}: Refactoring-aware Binary Adaptation of
  Evolving Libraries}. In \bibinfo{booktitle}{{\em Proceedings of the
  International Conference on Software Engineering (ICSE)}}.
\newblock


\bibitem[\protect\citeauthoryear{Egele, Brumley, Fratantonio, and
  Kruegel}{Egele et~al\mbox{.}}{2013}]%
        {Egele:CrytographicAndroidMisuse:CCS:2013}
\bibfield{author}{\bibinfo{person}{M. Egele}, \bibinfo{person}{D. Brumley},
  \bibinfo{person}{Y. Fratantonio}, {and} \bibinfo{person}{C. Kruegel}.}
  \bibinfo{year}{2013}\natexlab{}.
\newblock \showarticletitle{An Empirical Study of Cryptographic Misuse in
  {A}ndroid Applications}. In \bibinfo{booktitle}{{\em Proceedings of the ACM
  SIGSAC Conference on Computer \& Communications Security (CCS)}}.
\newblock


\bibitem[\protect\citeauthoryear{Elmalaki, Wanner, and Srivastava}{Elmalaki
  et~al\mbox{.}}{2015}]%
        {Elmalaki:CareDroid:MobiCom:2015}
\bibfield{author}{\bibinfo{person}{S. Elmalaki}, \bibinfo{person}{L. Wanner},
  {and} \bibinfo{person}{M. Srivastava}.} \bibinfo{year}{2015}\natexlab{}.
\newblock \showarticletitle{{CAreDroid}: Adaptation Framework for {A}ndroid
  Context-Aware Applications}. In \bibinfo{booktitle}{{\em Proceedings of the
  Annual International Conference on Mobile Computing and Networking
  (MobiCom)}}.
\newblock


\bibitem[\protect\citeauthoryear{Falcone, Currea, and Jaber}{Falcone
  et~al\mbox{.}}{2012}]%
        {Falcone:Runtime:2012}
\bibfield{author}{\bibinfo{person}{Y. Falcone}, \bibinfo{person}{S. Currea},
  {and} \bibinfo{person}{M. Jaber}.} \bibinfo{year}{2012}\natexlab{}.
\newblock \showarticletitle{Runtime Verification and Enforcement for {A}ndroid
  Applications with {RV-Droid}}. In \bibinfo{booktitle}{{\em Proceedings of the
  International Conference on Runtime Verification (RV)}}.
\newblock


\bibitem[\protect\citeauthoryear{Foundation}{Foundation}{2019a}]%
        {Acceleo}
\bibfield{author}{\bibinfo{person}{Eclipse Foundation}.}
  \bibinfo{year}{2019}\natexlab{a}.
\newblock \bibinfo{title}{{Acceleo}}.
\newblock \bibinfo{howpublished}{\url{https://www.eclipse.org/acceleo/}}.
  (\bibinfo{year}{2019}).
\newblock
\newblock
\shownote{[Online; accessed 21 January 2019].}


\bibitem[\protect\citeauthoryear{Foundation}{Foundation}{2019b}]%
        {Eclipse}
\bibfield{author}{\bibinfo{person}{Eclipse Foundation}.}
  \bibinfo{year}{2019}\natexlab{b}.
\newblock \bibinfo{title}{{Eclipse}}.
\newblock \bibinfo{howpublished}{\url{http://www.eclipse.org/}}.
  (\bibinfo{year}{2019}).
\newblock
\newblock
\shownote{[Online; accessed 21 January 2019].}


\bibitem[\protect\citeauthoryear{Foundation}{Foundation}{2019c}]%
        {EMF}
\bibfield{author}{\bibinfo{person}{Eclipse Foundation}.}
  \bibinfo{year}{2019}\natexlab{c}.
\newblock \bibinfo{title}{{Eclipse Modeling Framework}}.
\newblock \bibinfo{howpublished}{\url{https://www.eclipse.org/modeling/emf/}}.
   (\bibinfo{year}{2019}).
\newblock
\newblock
\shownote{[Online; accessed 21 January 2019].}


\bibitem[\protect\citeauthoryear{Foundation}{Foundation}{2019d}]%
        {GMF}
\bibfield{author}{\bibinfo{person}{Eclipse Foundation}.}
  \bibinfo{year}{2019}\natexlab{d}.
\newblock \bibinfo{title}{{Graphical Modeling Project}}.
\newblock \bibinfo{howpublished}{\url{http://www.eclipse.org/modeling/gmp/}}.
  (\bibinfo{year}{2019}).
\newblock
\newblock
\shownote{[Online; accessed 21 January 2019].}


\bibitem[\protect\citeauthoryear{Foundation}{Foundation}{2019e}]%
        {Appium_2017}
\bibfield{author}{\bibinfo{person}{JS Foundation}.}
  \bibinfo{year}{2019}\natexlab{e}.
\newblock \bibinfo{title}{{Appium}}.
\newblock \bibinfo{howpublished}{\url{http://appium.io}}.
  (\bibinfo{year}{2019}).
\newblock
\newblock
\shownote{[Online; accessed 4 January 2019].}


\bibitem[\protect\citeauthoryear{Gazzola, Micucci, and Mariani}{Gazzola
  et~al\mbox{.}}{2019}]%
        {Gazzola:Repair:TSE:2017}
\bibfield{author}{\bibinfo{person}{L. Gazzola}, \bibinfo{person}{D. Micucci},
  {and} \bibinfo{person}{L. Mariani}.} \bibinfo{year}{2019}\natexlab{}.
\newblock \showarticletitle{Automatic Software Repair: A Survey}.
\newblock \bibinfo{journal}{{\em IEEE Transactions on Software Engineering
  (TSE)\/}} \bibinfo{volume}{45}, \bibinfo{number}{1} (\bibinfo{year}{2019}),
  \bibinfo{pages}{34--67}.
\newblock
\showISSN{0098-5589}


\bibitem[\protect\citeauthoryear{Joorabchi, Mesbah, and Kruchten}{Joorabchi
  et~al\mbox{.}}{2013}]%
        {Joorabchi_RCM_2013}
\bibfield{author}{\bibinfo{person}{M.~E. Joorabchi}, \bibinfo{person}{A.
  Mesbah}, {and} \bibinfo{person}{P. Kruchten}.}
  \bibinfo{year}{2013}\natexlab{}.
\newblock \showarticletitle{Real Challenges in Mobile App Development}. In
  \bibinfo{booktitle}{{\em Proceedings of the International Symposium on
  Empirical Software Engineering and Measurement (ESEM)}}.
\newblock


\bibitem[\protect\citeauthoryear{Kelly and Tolvanen}{Kelly and
  Tolvanen}{2008}]%
        {Kelly:CodeGeneration:2008}
\bibfield{author}{\bibinfo{person}{S. Kelly} {and} \bibinfo{person}{J.-P.
  Tolvanen}.} \bibinfo{year}{2008}\natexlab{}.
\newblock \bibinfo{booktitle}{{\em Domain-Specific Modeling: Enabling Full Code
  Generation}}.
\newblock \bibinfo{publisher}{Wiley}.
\newblock


\bibitem[\protect\citeauthoryear{Khalid, Nagappan, Shihab, and Hassan}{Khalid
  et~al\mbox{.}}{2014}]%
        {Khalid_PDT_2014}
\bibfield{author}{\bibinfo{person}{H. Khalid}, \bibinfo{person}{M. Nagappan},
  \bibinfo{person}{E. Shihab}, {and} \bibinfo{person}{A.~E. Hassan}.}
  \bibinfo{year}{2014}\natexlab{}.
\newblock \showarticletitle{Prioritizing the Devices to Test Your App on: A
  Case Study of Android Game Apps}. In \bibinfo{booktitle}{{\em Proceedings of
  the Joint Meeting of the European Software Engineering Conference and the ACM
  SIGSOFT Symposium on the Foundations of Software Engineering (ESEC/FSE)}}.
\newblock


\bibitem[\protect\citeauthoryear{Kiczales, Lamping, Mendhekar, Maeda, Lopes,
  Loingtier, and Irwin}{Kiczales et~al\mbox{.}}{1997}]%
        {AOP97}
\bibfield{author}{\bibinfo{person}{G. Kiczales}, \bibinfo{person}{J. Lamping},
  \bibinfo{person}{A. Mendhekar}, \bibinfo{person}{C. Maeda},
  \bibinfo{person}{C. Lopes}, \bibinfo{person}{J. Loingtier}, {and}
  \bibinfo{person}{J. Irwin}.} \bibinfo{year}{1997}\natexlab{}.
\newblock \showarticletitle{Aspect-oriented programming}. In
  \bibinfo{booktitle}{{\em Proceedings of the European Conference on
  Object-Oriented Programming (ECOOP)}}.
\newblock


\bibitem[\protect\citeauthoryear{Kong, Cen, and Jin}{Kong
  et~al\mbox{.}}{2015}]%
        {Kong:AUTOREB:CCS:2015}
\bibfield{author}{\bibinfo{person}{D. Kong}, \bibinfo{person}{L. Cen}, {and}
  \bibinfo{person}{H. Jin}.} \bibinfo{year}{2015}\natexlab{}.
\newblock \showarticletitle{{AUTOREB}: Automatically Understanding the
  Review-to-Behavior Fidelity in {A}ndroid Applications}. In
  \bibinfo{booktitle}{{\em Proceedings of the ACM SIGSAC Conference on Computer
  and Communications Security (CCS)}}.
\newblock


\bibitem[\protect\citeauthoryear{Li and Halfond}{Li and Halfond}{2014}]%
        {Li:GreenProgramming:GREENS:2014}
\bibfield{author}{\bibinfo{person}{D. Li} {and} \bibinfo{person}{W.~G.~J.
  Halfond}.} \bibinfo{year}{2014}\natexlab{}.
\newblock \showarticletitle{An Investigation into Energy-saving Programming
  Practices for {A}ndroid Smartphone App Development}. In
  \bibinfo{booktitle}{{\em Proceedings of the International Workshop on Green
  and Sustainable Software (GREENS)}}.
\newblock


\bibitem[\protect\citeauthoryear{Li, Bissyand{\'e}, Octeau, and Klein}{Li
  et~al\mbox{.}}{2016}]%
        {Li:DroidRA:ISSTA:2016}
\bibfield{author}{\bibinfo{person}{L. Li}, \bibinfo{person}{T.~F.
  Bissyand{\'e}}, \bibinfo{person}{D. Octeau}, {and} \bibinfo{person}{J.
  Klein}.} \bibinfo{year}{2016}\natexlab{}.
\newblock \showarticletitle{{DroidRA}: Taming Reflection to Support
  Whole-program Analysis of {A}ndroid Apps}. In \bibinfo{booktitle}{{\em
  Proceedings of the International Symposium on Software Testing and Analysis
  (ISSTA)}}.
\newblock


\bibitem[\protect\citeauthoryear{Li and Zhou}{Li and Zhou}{2005}]%
        {Li:PRMiner:ESECFSE:2005}
\bibfield{author}{\bibinfo{person}{Z. Li} {and} \bibinfo{person}{Y. Zhou}.}
  \bibinfo{year}{2005}\natexlab{}.
\newblock \showarticletitle{{PR-Miner}: Automatically Extracting Implicit
  Programming Rules and Detecting Violations in Large Software Code}. In
  \bibinfo{booktitle}{{\em Proceedings of the European Software Engineering
  Conference held jointly with the ACM SIGSOFT International Symposium on
  Foundations of Software Engineering (ESEC/FSE)}}.
\newblock


\bibitem[\protect\citeauthoryear{Ligatti, Bauer, and Walker}{Ligatti
  et~al\mbox{.}}{2005}]%
        {Ligatti:Edit:JIS:2005}
\bibfield{author}{\bibinfo{person}{J. Ligatti}, \bibinfo{person}{L. Bauer},
  {and} \bibinfo{person}{D. Walker}.} \bibinfo{year}{2005}\natexlab{}.
\newblock \showarticletitle{Edit automata: enforcement mechanisms for run-time
  security policies}.
\newblock \bibinfo{journal}{{\em International Journal of Information
  Security\/}} \bibinfo{volume}{4}, \bibinfo{number}{1} (\bibinfo{year}{2005}),
  \bibinfo{pages}{2--16}.
\newblock


\bibitem[\protect\citeauthoryear{Linares-V\'{a}squez, Bavota,
  Bernal-C\'{a}rdenas, Di~Penta, Oliveto, and Poshyvanyk}{Linares-V\'{a}squez
  et~al\mbox{.}}{2013a}]%
        {LinaresVasquez_ACF_2013}
\bibfield{author}{\bibinfo{person}{M. Linares-V\'{a}squez}, \bibinfo{person}{G.
  Bavota}, \bibinfo{person}{C. Bernal-C\'{a}rdenas}, \bibinfo{person}{M.
  Di~Penta}, \bibinfo{person}{R. Oliveto}, {and} \bibinfo{person}{D.
  Poshyvanyk}.} \bibinfo{year}{2013}\natexlab{a}.
\newblock \showarticletitle{API Change and Fault Proneness: A Threat to the
  Success of Android Apps}. In \bibinfo{booktitle}{{\em Proceedings of the
  Joint Meeting of the European Software Engineering Conference and the ACM
  SIGSOFT Symposium on the Foundations of Software Engineering (ESEC/FSE)}}.
\newblock


\bibitem[\protect\citeauthoryear{Linares-V\'{a}squez, Bavota,
  Bernal-C\'{a}rdenas, Penta, Oliveto, and Poshyvanyk}{Linares-V\'{a}squez
  et~al\mbox{.}}{2013b}]%
        {Linares-Vasquez:AndoirdAPIChange:ESECFSE:2013}
\bibfield{author}{\bibinfo{person}{M. Linares-V\'{a}squez}, \bibinfo{person}{G.
  Bavota}, \bibinfo{person}{C. Bernal-C\'{a}rdenas}, \bibinfo{person}{M.~Di
  Penta}, \bibinfo{person}{R. Oliveto}, {and} \bibinfo{person}{D. Poshyvanyk}.}
  \bibinfo{year}{2013}\natexlab{b}.
\newblock \showarticletitle{{API} Change and Fault Proneness: A Threat to the
  Success of {A}ndroid Apps}. In \bibinfo{booktitle}{{\em Proceedings of the
  Joint Meeting of the European Software Engineering Conference and the ACM
  SIGSOFT Symposium on the Foundations of Software Engineering (ESEC/FSE)}}.
\newblock


\bibitem[\protect\citeauthoryear{Liu, Wu, Yan, and Zhang}{Liu
  et~al\mbox{.}}{2016}]%
        {Liu:ResourceLeaks:ISSRE:2016}
\bibfield{author}{\bibinfo{person}{J. Liu}, \bibinfo{person}{T. Wu},
  \bibinfo{person}{J. Yan}, {and} \bibinfo{person}{J. Zhang}.}
  \bibinfo{year}{2016}\natexlab{}.
\newblock \showarticletitle{Fixing Resource Leaks in {A}ndroid Apps with
  Light-Weight Static Analysis and Low-Overhead Instrumentation}. In
  \bibinfo{booktitle}{{\em Proceedings of the International Symposium on
  Software Reliability Engineering (ISSRE)}}.
\newblock


\bibitem[\protect\citeauthoryear{Magalh\~{a}es and Silva}{Magalh\~{a}es and
  Silva}{2015}]%
        {Magalhaes_SSH_2015}
\bibfield{author}{\bibinfo{person}{J.~P. Magalh\~{a}es} {and}
  \bibinfo{person}{L.~M. Silva}.} \bibinfo{year}{2015}\natexlab{}.
\newblock \showarticletitle{{SH\~{O}WA}: A Self-Healing Framework for Web-Based
  Applications}.
\newblock \bibinfo{journal}{{\em ACM Transactions on Autonomous and Adaptive
  Systems\/}} \bibinfo{volume}{10}, \bibinfo{number}{1} (\bibinfo{year}{2015}),
  \bibinfo{pages}{4:1--4:28}.
\newblock


\bibitem[\protect\citeauthoryear{Mancinelli and Inverardi}{Mancinelli and
  Inverardi}{2006}]%
        {Mancinelli:ResourceModel:SEAMS:2006}
\bibfield{author}{\bibinfo{person}{F. Mancinelli} {and} \bibinfo{person}{P.
  Inverardi}.} \bibinfo{year}{2006}\natexlab{}.
\newblock \showarticletitle{A Resource Model for Adaptable Applications}. In
  \bibinfo{booktitle}{{\em Proceedings of the International Workshop on
  Self-adaptation and Self-managing Systems (SEAMS)}}.
\newblock


\bibitem[\protect\citeauthoryear{Mariani, Pastore, and Pezz{\`e}}{Mariani
  et~al\mbox{.}}{2011}]%
        {Mariani:BCT:TSE:2011}
\bibfield{author}{\bibinfo{person}{L. Mariani}, \bibinfo{person}{F. Pastore},
  {and} \bibinfo{person}{M. Pezz{\`e}}.} \bibinfo{year}{2011}\natexlab{}.
\newblock \showarticletitle{Dynamic Analysis for Diagnosing Integration
  Faults}.
\newblock \bibinfo{journal}{{\em IEEE Transactions on Software Engineering
  (TSE)\/}} \bibinfo{volume}{37}, \bibinfo{number}{4} (\bibinfo{year}{2011}),
  \bibinfo{pages}{486--508}.
\newblock


\bibitem[\protect\citeauthoryear{McDonnell, Ray, and Kim}{McDonnell
  et~al\mbox{.}}{2013}]%
        {McDonnell_ESA_2013}
\bibfield{author}{\bibinfo{person}{T. McDonnell}, \bibinfo{person}{B. Ray},
  {and} \bibinfo{person}{M. Kim}.} \bibinfo{year}{2013}\natexlab{}.
\newblock \showarticletitle{An Empirical Study of API Stability and Adoption in
  the Android Ecosystem}. In \bibinfo{booktitle}{{\em Proceedings of the
  International Conference on Software Maintenance (ICSM)}}.
\newblock


\bibitem[\protect\citeauthoryear{Muccini, Francesco, and Esposito}{Muccini
  et~al\mbox{.}}{2012}]%
        {Muccini_STM_2012}
\bibfield{author}{\bibinfo{person}{H. Muccini}, \bibinfo{person}{A.~Di
  Francesco}, {and} \bibinfo{person}{P. Esposito}.}
  \bibinfo{year}{2012}\natexlab{}.
\newblock \showarticletitle{Software testing of mobile applications: Challenges
  and future research directions}. In \bibinfo{booktitle}{{\em Proceedings of
  the International Workshop on Automation of Software Test (AST)}}.
\newblock


\bibitem[\protect\citeauthoryear{Mulliner, Oberheide, Robertson, and
  Kirda}{Mulliner et~al\mbox{.}}{2013}]%
        {Mulliner_PST_2013}
\bibfield{author}{\bibinfo{person}{C. Mulliner}, \bibinfo{person}{J.
  Oberheide}, \bibinfo{person}{W. Robertson}, {and} \bibinfo{person}{E.
  Kirda}.} \bibinfo{year}{2013}\natexlab{}.
\newblock \showarticletitle{{PatchDroid}: Scalable Third-party Security Patches
  for {A}ndroid Devices}. In \bibinfo{booktitle}{{\em Proceedings of the Annual
  Computer Security Applications Conference (ACSAC)}}.
\newblock


\bibitem[\protect\citeauthoryear{Neisse, Steri, Geneiatakis, and Fovino}{Neisse
  et~al\mbox{.}}{2016}]%
        {Neisse:EnforcingPrivacyAndroid:ComputerSecurity:2016}
\bibfield{author}{\bibinfo{person}{R. Neisse}, \bibinfo{person}{G. Steri},
  \bibinfo{person}{D. Geneiatakis}, {and} \bibinfo{person}{I.~N. Fovino}.}
  \bibinfo{year}{2016}\natexlab{}.
\newblock \showarticletitle{A privacy enforcing framework for Android
  applications}.
\newblock \bibinfo{journal}{{\em Computers \& Security\/}}
  \bibinfo{volume}{62} (\bibinfo{year}{2016}), \bibinfo{pages}{257--277}.
\newblock


\bibitem[\protect\citeauthoryear{OMG}{OMG}{2012}]%
        {OCL}
\bibfield{author}{\bibinfo{person}{OMG}.} \bibinfo{year}{2012}\natexlab{}.
\newblock \bibinfo{title}{{Object Constraint Language}}.
\newblock
  \bibinfo{howpublished}{\url{https://modeling-languages.com/ocl-tutorial/}}.
  (\bibinfo{year}{2012}).
\newblock
\newblock
\shownote{[Online; accessed 21 January 2019].}


\bibitem[\protect\citeauthoryear{Riganelli, Micucci, and Mariani}{Riganelli
  et~al\mbox{.}}{2016}]%
        {Riganelli:HealingDataLos:IWSF:2016}
\bibfield{author}{\bibinfo{person}{O. Riganelli}, \bibinfo{person}{D. Micucci},
  {and} \bibinfo{person}{L. Mariani}.} \bibinfo{year}{2016}\natexlab{}.
\newblock \showarticletitle{{Healing Data Loss Problems in {A}ndroid Apps}}. In
  \bibinfo{booktitle}{{\em Proceedings of the International Workshop on
  Software Faults (IWSF), co-located with the International Symposium on
  Software Reliability Engineering (ISSRE)}}.
\newblock


\bibitem[\protect\citeauthoryear{Riganelli, Micucci, and Mariani}{Riganelli
  et~al\mbox{.}}{2017a}]%
        {Riganelli:ProactiveLibraries:SEAMS:2017}
\bibfield{author}{\bibinfo{person}{O. Riganelli}, \bibinfo{person}{D. Micucci},
  {and} \bibinfo{person}{L. Mariani}.} \bibinfo{year}{2017}\natexlab{a}.
\newblock \showarticletitle{Policy enforcement with proactive libraries}. In
  \bibinfo{booktitle}{{\em Proceedings of the International Symposium on
  Software Engineering for Adaptive and Self-Managing Systems (SEAMS)}}.
\newblock


\bibitem[\protect\citeauthoryear{Riganelli, Micucci, and Mariani}{Riganelli
  et~al\mbox{.}}{2018}]%
        {Riganelli:EnforcerReusability:ISOLA:2018}
\bibfield{author}{\bibinfo{person}{O. Riganelli}, \bibinfo{person}{D. Micucci},
  {and} \bibinfo{person}{L. Mariani}.} \bibinfo{year}{2018}\natexlab{}.
\newblock \showarticletitle{Increasing the Reusability of Enforcers with
  Lifecycle Events}. In \bibinfo{booktitle}{{\em Proceedings of the
  International Symposium on Leveraging Applications of Formal Methods,
  Verification and Validation (ISOLA)}}.
\newblock


\bibitem[\protect\citeauthoryear{Riganelli, Micucci, and Mariani}{Riganelli
  et~al\mbox{.}}{2019a}]%
        {Riganelli:SPE:ResourceLeak:2019}
\bibfield{author}{\bibinfo{person}{O. Riganelli}, \bibinfo{person}{D. Micucci},
  {and} \bibinfo{person}{L. Mariani}.} \bibinfo{year}{2019}\natexlab{a}.
\newblock \showarticletitle{From source code to test cases: A comprehensive
  benchmark for resource leak detection in Android apps}.
\newblock \bibinfo{journal}{{\em Software: Practice and Experience\/}}
  \bibinfo{volume}{49}, \bibinfo{number}{3} (\bibinfo{year}{2019}),
  \bibinfo{pages}{540--548}.
\newblock


\bibitem[\protect\citeauthoryear{Riganelli, Micucci, Mariani, and
  Falcone}{Riganelli et~al\mbox{.}}{2017b}]%
        {RV17}
\bibfield{author}{\bibinfo{person}{O. Riganelli}, \bibinfo{person}{D. Micucci},
  \bibinfo{person}{L. Mariani}, {and} \bibinfo{person}{Y. Falcone}.}
  \bibinfo{year}{2017}\natexlab{b}.
\newblock \showarticletitle{Verifying Policy Enforcers}. In
  \bibinfo{booktitle}{{\em Proceedings of the International Conference on
  Runtime Verification (RV)}}.
\newblock


\bibitem[\protect\citeauthoryear{Riganelli, Mobilio, Micucci, and
  Mariani}{Riganelli et~al\mbox{.}}{2019b}]%
        {Riganelli:MSR:DLBenchmark}
\bibfield{author}{\bibinfo{person}{O. Riganelli}, \bibinfo{person}{M. Mobilio},
  \bibinfo{person}{D. Micucci}, {and} \bibinfo{person}{L. Mariani}.}
  \bibinfo{year}{2019}\natexlab{b}.
\newblock \showarticletitle{A Benchmark of Data Loss Bugs for Android Apps}. In
  \bibinfo{booktitle}{{\em Proceedings of the International Conference on
  Mining Software Repositories (MSR)}}.
\newblock


\bibitem[\protect\citeauthoryear{Rouvoy, Beauvois, Lozano, Lorenzo, and
  Eliassen}{Rouvoy et~al\mbox{.}}{2008}]%
        {Rouvoy:MUSIC:MobMid:2008}
\bibfield{author}{\bibinfo{person}{R. Rouvoy}, \bibinfo{person}{M. Beauvois},
  \bibinfo{person}{L. Lozano}, \bibinfo{person}{J. Lorenzo}, {and}
  \bibinfo{person}{F. Eliassen}.} \bibinfo{year}{2008}\natexlab{}.
\newblock \showarticletitle{{MUSIC}: An Autonomous Platform Supporting
  Self-adaptive Mobile Applications}. In \bibinfo{booktitle}{{\em Proceedings
  of the Workshop on Mobile Middleware: Embracing the Personal Communication
  Device (MobMid)}}.
\newblock


\bibitem[\protect\citeauthoryear{Seiger, Huber, and Schlegel}{Seiger
  et~al\mbox{.}}{2018}]%
        {Seiger2016}
\bibfield{author}{\bibinfo{person}{Ronny Seiger}, \bibinfo{person}{Steffen
  Huber}, {and} \bibinfo{person}{Thomas Schlegel}.}
  \bibinfo{year}{2018}\natexlab{}.
\newblock \showarticletitle{Toward an execution system for self-healing
  workflows in cyber-physical systems}.
\newblock \bibinfo{journal}{{\em Software {\&} Systems Modeling\/}}
  \bibinfo{volume}{17}, \bibinfo{number}{2} (\bibinfo{year}{2018}),
  \bibinfo{pages}{551--572}.
\newblock
\showISSN{1619-1374}


\bibitem[\protect\citeauthoryear{Shan, Azim, and Neamtiu}{Shan
  et~al\mbox{.}}{2016}]%
        {Shan:ResumerRestartErrors:OOPSLA:2016}
\bibfield{author}{\bibinfo{person}{Z. Shan}, \bibinfo{person}{T. Azim}, {and}
  \bibinfo{person}{I. Neamtiu}.} \bibinfo{year}{2016}\natexlab{}.
\newblock \showarticletitle{Finding Resume and Restart Errors in {A}ndroid
  Applications}. In \bibinfo{booktitle}{{\em Proceedings of the ACM SIGPLAN
  International Conference on Object-Oriented Programming, Systems, Languages,
  and Applications (OOPSLA)}}.
\newblock


\bibitem[\protect\citeauthoryear{Stahl and Voelter}{Stahl and Voelter}{2006}]%
        {Stahl:MDSD:2006}
\bibfield{author}{\bibinfo{person}{T. Stahl} {and} \bibinfo{person}{M.
  Voelter}.} \bibinfo{year}{2006}\natexlab{}.
\newblock \bibinfo{booktitle}{{\em Model-Driven Software Development:
  Technology, Engineering, Management}}.
\newblock \bibinfo{publisher}{John Wiley \& Sons}.
\newblock


\bibitem[\protect\citeauthoryear{Wasylkowski and Zeller}{Wasylkowski and
  Zeller}{2009}]%
        {Wasylkowski:MiningTemporalSpec:ASE:2009}
\bibfield{author}{\bibinfo{person}{A. Wasylkowski} {and} \bibinfo{person}{A.
  Zeller}.} \bibinfo{year}{2009}\natexlab{}.
\newblock \showarticletitle{Mining Temporal Specifications from Object Usage}.
  In \bibinfo{booktitle}{{\em Proceedings of the IEEE/ACM International
  Conference on Automated Software Engineering (ASE)}}.
\newblock


\bibitem[\protect\citeauthoryear{Wei, Liu, and Cheung}{Wei
  et~al\mbox{.}}{2016}]%
        {Wei:AndroidFragmentation:ASE:2016}
\bibfield{author}{\bibinfo{person}{L. Wei}, \bibinfo{person}{Y. Liu}, {and}
  \bibinfo{person}{S.-C. Cheung}.} \bibinfo{year}{2016}\natexlab{}.
\newblock \showarticletitle{Taming {A}ndroid Fragmentation: Characterizing and
  Detecting Compatibility Issues for {A}ndroid Apps}. In
  \bibinfo{booktitle}{{\em Proceedings of the IEEE/ACM International Conference
  on Automated Software Engineering (ASE)}}.
\newblock


\bibitem[\protect\citeauthoryear{Wu, Liu, Xu, Guo, Zhang, Yan, and Zhang}{Wu
  et~al\mbox{.}}{2016}]%
        {Wu:Callback:TSE:2016}
\bibfield{author}{\bibinfo{person}{T. Wu}, \bibinfo{person}{J. Liu},
  \bibinfo{person}{Z. Xu}, \bibinfo{person}{C. Guo}, \bibinfo{person}{Y.
  Zhang}, \bibinfo{person}{J. Yan}, {and} \bibinfo{person}{J. Zhang}.}
  \bibinfo{year}{2016}\natexlab{}.
\newblock \showarticletitle{Light-Weight, Inter-Procedural and Callback-Aware
  Resource Leak Detection for {A}ndroid Apps}.
\newblock \bibinfo{journal}{{\em IEEE Transactions on Software Engineering
  (TSE)\/}} \bibinfo{volume}{42}, \bibinfo{number}{11} (\bibinfo{year}{2016}),
  \bibinfo{pages}{1054--1076}.
\newblock


\bibitem[\protect\citeauthoryear{XDA}{XDA}{2019}]%
        {Xposed_2016}
\bibfield{author}{\bibinfo{person}{XDA}.} \bibinfo{year}{2019}\natexlab{}.
\newblock \bibinfo{title}{{Xposed}}.
\newblock \bibinfo{howpublished}{\url{http://repo.xposed.info/}}.
  (\bibinfo{year}{2019}).
\newblock
\newblock
\shownote{[Online; accessed 21 January 2019].}


\bibitem[\protect\citeauthoryear{Zhang and Yin}{Zhang and Yin}{2014}]%
        {Zhang_AAG_2014}
\bibfield{author}{\bibinfo{person}{M. Zhang} {and} \bibinfo{person}{H. Yin}.}
  \bibinfo{year}{2014}\natexlab{}.
\newblock \showarticletitle{AppSealer: Automatic Generation of
  Vulnerability-Specific Patches for Preventing Component Hijacking Attacks in
  {A}ndroid Applications}. In \bibinfo{booktitle}{{\em Proceedings of the
  Annual Network and Distributed System Security Symposium (NDSS)}}.
\newblock


\end{thebibliography}

\end{document}